\pdfoutput=1

\documentclass[11pt]{article}

\usepackage{placeins}
 \usepackage{float}
\usepackage{acl}

\usepackage{times}
\usepackage{latexsym}

\usepackage[T1]{fontenc}

\usepackage[utf8]{inputenc}

\usepackage{microtype}

\usepackage{inconsolata}

\usepackage{graphicx}
\usepackage{amsmath}
\usepackage{xcolor}
\usepackage[normalem]{ulem}

\usepackage{tikz}
\usetikzlibrary{shapes,arrows}

\usepackage{tikz}
\usepackage[most]{tcolorbox}
\usepackage{fontawesome5}
\usepackage{booktabs}

\usepackage{fontawesome5}

\definecolor{lightgray}{RGB}{245,245,245}
\definecolor{lightblue}{RGB}{230,240,255}

\newcommand{\chatbox}[4]{
    \begin{tcolorbox}[
        enhanced,
        boxrule=0.5pt,
        colback=#1,
        colframe=#1,
        arc=10pt,
        outer arc=10pt,
        leftrule=0pt,
        rightrule=0pt,
        toprule=0pt,
        bottomrule=0pt,
        left=#2,
        right=#3,
        boxsep=0pt,
        top=2pt,
        bottom=2pt,
    ]
        #4
    \end{tcolorbox}
}

\author{
  \textbf{Alexander Wuttke\textsuperscript{1}},
  \textbf{Matthias Aßenmacher\textsuperscript{1,2}},
 \textbf{Christopher Klamm\textsuperscript{2}},
 \\
  \textbf{Max M. Lang\textsuperscript{3}},
  \textbf{Quirin Würschinger\textsuperscript{1}},
  \textbf{Frauke Kreuter\textsuperscript{1,2}}
\\
\\
  \textsuperscript{1}LMU Munich,
  \textsuperscript{2}Munich Center for Machine Learning (MCML),\\
  \textsuperscript{3}University of Mannheim,
  \textsuperscript{4}University of Oxford
\\
  \small{
    \textbf{Correspondence:} \href{mailto:a.wuttke@lmu.de}{a.wuttke@lmu.de}
  }
}

\title{AI Conversational Interviewing:\\ Transforming Surveys with LLMs as Adaptive Interviewers}

\begin{document}

\newcommand{\wordcountHuman}[0]{32.81 }
\newcommand{\wordcountAI}[0]{52.39 }
\newcommand{\readbilityaiall}[0]{77.66 }
\newcommand{\readbilityaiwvoice}[0]{48.32 }
\newcommand{\readbilityhumanall}[0]{62.22 }


\newcommand{\CK}[3][red]{{\color{#1}\sout{#2} \textbf{CK:} #3}}
\newcommand{\FK}[3][red]{{\color{#1}\sout{#2} \textbf{FK:} #3}}
\newcommand{\AW}[3][red]{{\color{#1}\sout{#2} \textbf{AW:} #3}}
\newcommand{\ML}[3][red]{{\color{#1}\sout{#2} \textbf{ML:} #3}}
\newcommand{\MA}[3][red]{{\color{#1}\sout{#2} \textbf{MA:} #3}}
\newcommand{\QW}[3][red]{{\color{#1}\sout{#2} \textbf{QW:} #3}}

\maketitle

\begin{abstract}
Traditional methods for eliciting people's opinions face a trade-off between depth and scale: structured surveys enable large-scale data collection but limit respondents' ability to voice their opinions in their own words, while conversational interviews provide deeper insights but are resource-intensive. This study explores the potential of replacing human interviewers with large language models (LLMs) to conduct scalable conversational interviews. Our goal is to assess the performance of AI Conversational Interviewing and to identify opportunities for improvement in a controlled environment. We conducted a small-scale, in-depth study with university students who were randomly assigned to a conversational interview by either AI or human interviewers, both employing identical questionnaires on political topics. Various quantitative and qualitative measures assessed interviewer adherence to guidelines, response quality, participant engagement, and overall interview efficacy. The findings indicate the viability of AI Conversational Interviewing in producing quality data comparable to traditional methods, with the added benefit of scalability. We publish our data and materials for re-use and present specific recommendations for effective implementation. \end{abstract}

\section{Introduction}
\label{intro}
Structured surveys are popular tools to assess public opinion  \citep{groves_survey_2009, kertzer2022experiments, stantcheva2023run}. These surveys typically gather individual orientations through self-reports, asking respondents to select from predefined options on fixed questions. This method allows for efficient data collection across large populations, producing structured, tabular data that is straightforward to analyze and comparable across respondents \citep{krosnick1999survey, groves_survey_2009}. Due to these benefits, structured surveys hold a prominent position in both academic and commercial research.

Despite their established utility, structured surveys with predefined response options have significant limitations \citep{schwarz1987response,kash2013open}. Their static and impersonal nature often leads to respondent fatigue, which can diminish engagement and, consequently, the quality of responses \citep{krosnick1999survey, jeong2023exhaustive}. More critically, the rigid format of these surveys constrains respondents from fully expressing their thoughts, restricting them from offering responses that researchers may not have anticipated \citep{chang2021ambivalence,esses2002expanding,reja2003open,baburajan2022open, duck2023ends}. 

\begin{figure*}[!ht]
    \centering
    \includegraphics[width=.8\textwidth]{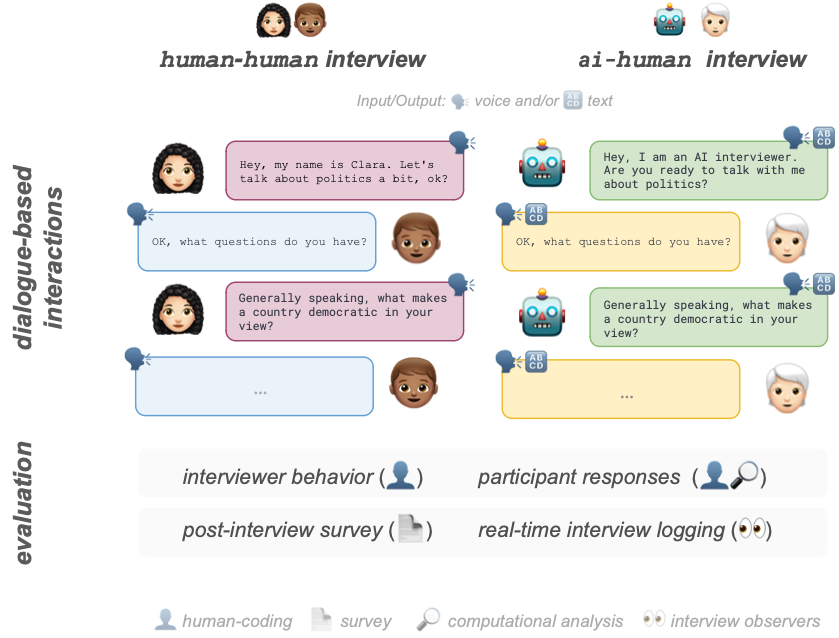}
    \caption{Illustration of the concurrent interview settings (human- vs. AI-conducted) and the various metrics (\faIcon{user}, \faIcon{eye}, \faIcon{file-alt} and \faIcon{search}) applied to assess interview quality.}
    \label{fig:ai_human_setting}
\end{figure*}

This limitation hampers the discovery of new phenomena and prevents a comprehensive understanding of the full spectrum of people's attitudes.
An alternative to structured surveys is conversational interviewing, sometimes called in-depth or semi-structured or qualitative interviewing \citep{adeoye2021research, kallio2016systematic, newcomer_conducting_2015}. It involves interviewers engaging with respondents in a more open-ended format, allowing them to freely express their thoughts on topics of interest. The dynamic nature of conversational interviews helps alleviate respondent fatigue and permits the exploration of opinions beyond predefined response options. However, this approach requires skilled interviewers capable of conducting nuanced conversations, which limits its application to small sample sizes due to the associated costs.

So, survey research faces a trade-off between depth and scale: researchers must choose between conducting in-depth explorations with small groups through or large-scale but rigid surveys. However, recent advances in natural language processing \citep{dubey2024llama3herdmodels,üstün2024ayamodelinstructionfinetuned, workshop2023bloom176bparameteropenaccessmultilingual, costello2024durably} present new possibilities for addressing this dilemma. The conversational capabilities of instruction-finetuned large language models  \citep{wei2022finetunedlanguagemodelszeroshot,ouyang2022training} have made them applicable across various academic and industrial domains. Because LLMs can engage in human-like conversations \citep{cai2024largelanguagemodelsresemble, di2023pragmatic, palmer2023large}, they have the potential to assist or even replace human interviewers in conducting conversational interviews. By eliminating the costly need for human interviewers, LLMs could enable scalable in-depth conversations, potentially resolving the trade-off between depth and scale.

\paragraph{Contributions} 
We contribute to the emerging paradigm of AI Conversational Interviewing by conducting the first close-up investigation of its practical implementation and performance~(cf. Figure \ref{fig:ai_human_setting}):

\begin{itemize}
\item We provide a new comprehensive assessment pipeline of AI performance in conducting conversational interviews
\item We document the practical challenges participants face when interacting with an AI interviewer
\item We are the first to explore the performance of voice-assisted LLM-based interviewing
\item We are the first to perform a detailed comparative analysis of AI-conducted versus human-conducted conversational interviews
\item We pre-registered the study to ensure transparency in the research process
\item We publish code and data for reuse: \href{https://github.com/AIinterviewing/ai-conversational-interviewing-LaTeCH-CLfL2025}{https://github.com/AIinterviewing/ai-conversational-interviewing-LaTeCH-CLfL2025} 
\end{itemize}

\section{Related Work}

To implement and evaluate AI Conversational Interviews this study combines insights from three distinct lines of work that have rarely been combined.

\textbf{Advances in AI research} have facilitated multiple ongoing commercial and academic projects that use LLM-powered chatbots for in-depth, qualitative, or semi-structured interviews, as they are interchangeably called 
\citep{chopra_conducting_2023,weidmann_dialing_2024}. Although implementations vary, the studies collectively highlight the potential of LLMs for conducting conversational interviews. Yet, critical questions regarding the implementation remain unresolved and little is known about the relative performance compared with human-led interviews.

\textbf{Qualitative studies} have extensively explored best practices for conducting in-person interviews \citep{newcomer_conducting_2015}. Our approach is to build on these insights when implementing AI Conversational Interviewing.

\textbf{Studies in survey methodology} have extensively examined how different interview implementations influence responses. One line of research has focused on interviewer and mode effects \citep{mittereder_interviewerrespondent_2018, malhotra2007effect}. The presence of an interviewer significantly impacts respondents, often leading to greater engagement but also increasing the likelihood of socially desirable responses \citep{atkeson_nonresponse_2014, west_explaining_2016-1}. In this vein, studies on conversational interviewing has shown that a more active and flexible interviewer who engages with questions from respondents can improve data quality  \citep{schober1997does, davis2024ounce, mittereder_interviewerrespondent_2018}.\footnote{Our method is similar to traditional "conversational interviewing" in that it enhances flexibility during the interview. However, AI Conversational Interviewing differs by highlighting the flexibility of the respondents rather than the interviewer.} Another important factor is the input mode. Responses to open-ended questions vary depending on whether they are submitted via text or speech. Text input typically requires more effort, which can result in shorter but more carefully considered responses \citep{gavras_innovating_2022, hohne_sound_2024}. So, the responses will not necessarily be better or worse depending on input mode, but they will differ predictably, as text- and speech-based interviews elicit distinct psychological reactions from participants \citep{gavras_innovating_2022}.

\section{Study Design and Implementation}\label{sec:design}
Our study pursues two goals: (a) Assess the performance of AI Conversational Interviewing (in comparison to human-led interviewing) and (b) Identify problems and opportunities for improvement of AI Conversational Interviewing.

We conducted a small-N study among university students in a controlled environment. Ahead of data collection, we pre-registered our research questions, research design, and evaluation metrics (cf. \href{https://osf.io/c8ymh/}{OSF Registry}).

We conducted both AI-led and human-led interviews as part of a class activity, where students were randomly assigned to serve as either interviewers or respondents in the respective conditions. Identical questionnaires were used in both interview settings. After the interview, respondents filled out a structured questionnaire to evaluate their interview experience. In the AI interview condition, students monitored the interviewees in real-time to identify any technical issues.

\begin{figure*}[!ht]
\centering
\scalebox{0.9}{
\centering
\begin{tikzpicture}
    \node[inner sep=0pt] (background) at (0,0) {
        \begin{tikzpicture}
            
            
            \node[anchor=north west, inner sep=0] at (-8.5,3.3) {
                \chatbox{lightblue}{0.5cm}{0.5cm}{
                    \parbox{14cm}{\textbf{AI:} Let's talk about democracy. What good things come to mind about democracy in Western countries like Germany?\\
                    \scriptsize\textcolor{gray}{10:30 AM}}\\
                }
            };
            \node[anchor=north west, font=\tiny] at (-9.0,3.3) {\faIcon[solid]{robot}};
            
            \node[anchor=north west, inner sep=0] at (-6.5,1.5) {
                \chatbox{white}{0.5cm}{1.5cm}{
                    \textbf{Respondent:} A good thing is having a democracy that's less corrupt than other countries, like the recent election in Russia.\\
                    \scriptsize\textcolor{gray}{10:31 AM}\\
                }
            };
            \node[anchor=north east, font=\tiny] at (8.5,1.5) {\faIcon{user}};
            
            \node[anchor=north west, inner sep=0] at (-8.5,-0.3) {
                \chatbox{lightblue}{0.5cm}{-2.5cm}{
                    \parbox{14cm}{\textbf{AI:} You appreciate the less corrupt system in Western democracies, contrasting with less transparent elections elsewhere. What bad things come to mind about democracy in the West?\\
                    \scriptsize\textcolor{gray}{10:32 AM}}
                }
            };
            \node[anchor=north west, font=\tiny] at (-9.0,-0.3) {\faIcon[solid]{robot}};
            
            \node[anchor=north west] at (-5.5,-2.7) {
                \begin{tikzpicture}
                    \foreach \x in {0,0.3,0.6} {
                        \fill[lightblue] (\x,0) circle (0.05);
                    }
                \end{tikzpicture}
            };
            
            \draw[rounded corners] (-5.8,-3.5) rectangle (5.8,-3);
            \node[anchor=west, font=\scriptsize] at (-5.6,-3.25) {Type a message...};
            \node[anchor=east, font=\tiny] at (5.6,-3.25) {\faIcon{paper-plane}};
            \node[anchor=east, font=\tiny] at (5.2,-3.25) {\faIcon{microphone}};
        \end{tikzpicture}
    };
\end{tikzpicture}
}
\centering
\caption{Illustrative example of our used Chat Interface structure (with an interaction between an AI agent \faIcon{robot} and a user \faIcon{user}) of the AI in-depth interview, showcasing how the interviewer engages in \textit{active listening} by occasionally rehearsing the preceding answer, as instructed (cf. Appendix \ref{text:ai-interviewer-prompt}). The input field includes options for text input (\faIcon{paper-plane}) and voice input (\faIcon{microphone}).}
\label{fig:ai-interface-active-listening}
\end{figure*}
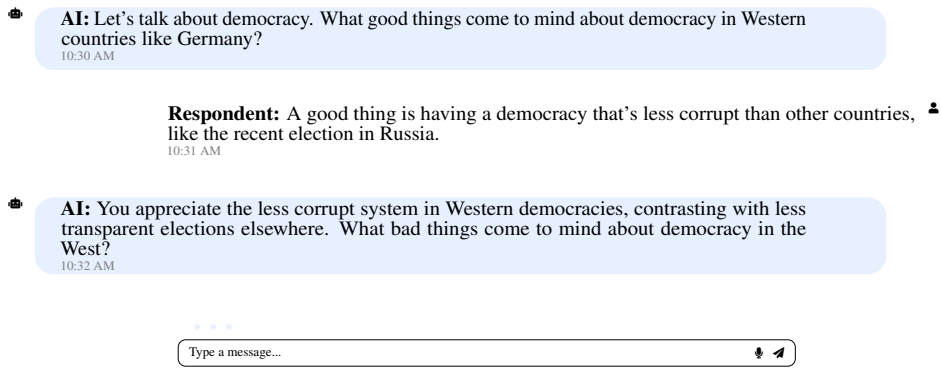

\subsection{Procedure}
The study was embedded in a student seminar on survey methodology that was hosted via Zoom. Students were informed that they would participate in a pilot study of conversational interviewing. The seminar proceeded with a detailed script (cf. Appendix \ref{text:guidelines-asking-questions}), lasting about 120 minutes:

\begin{enumerate}
    \item Participants were informed about the upcoming procedure, the technical requirements were laid out, and they were asked for consent to participate and collect their data.
    \item As preparation for the upcoming tasks, an instructor gave a 10-minute presentation about scientific approaches to interview respondents, and rules for good interviewer behavior.
    \item Students were paired up and randomly assigned different roles:
    \begin{enumerate}
        \item Students participated in both a human-conducted and an AI-conducted interview, with the sequence randomly assigned
        \item In the human-conducted interviews, students took on roles as either respondents or interviewers
        \item In the AI-conducted interviews, students served as either respondents or observers, monitoring for any technical issues during the interview
    \end{enumerate}
\end{enumerate}

\subsection{Model setup}
We implemented a voice-assisted AI Conversational Interviewing pipeline with GPT-4\footnote{GPT-4 turbo, version: 04/2024} and a Chainlit-based user interface, using the following task-adapted prompts (cf. Appendix \ref{text:ai-interviewer-prompt}):

\begin{enumerate}
    \item The \textbf{system} instruction to act as an interviewer (\textit{You are a survey interviewer named 'InterviewGPT', an AI interviewer, wanting to find out more about people's views [...]})
    \item the \textbf{user} instructions with specific guidelines, derived from the qualitative literature on human in-depth interviewing \citep{newcomer_conducting_2015}, specifying desirable and undesirable interviewer behavior (\textit{[...] Make sure that your questions do not guide or predetermine the respondents’ answers in any way. Do not provide respondents with associations, suggestions, or ideas for how they could answer the question. [...]})
    \item a \textbf{task} questionnaire on politics and democracy, developed by a democracy researcher among the authors (e.g. \textit{And what do you think “politics” is? How would you define this term?})
\end{enumerate}

\subsection{User interface}
To enable voice-assisted interviewing, we developed a user interface based on Chainlit\footnote{\url{https://chainlit.io/}},
with customization for audio input and output as shown in Figure~\ref{fig:ai-interface-active-listening}). Our voice-assisted implementation allowed respondents to choose between voice and text modes for both the model output (interviewer questions) and their input (responses). When respondents selected audio input, their speech was transcribed into text, which they could then review and edit before submitting their responses. This approach sought to blend the spontaneity and expressiveness of audio input with the precision and control offered by text-based refinements. For audio output, interviewer questions were displayed as text and could be delivered as voice upon the user’s request. We utilized OpenAI Whisper \citep{Radford2023Robust} for text-to-speech transcriptions of model-generated text.

\subsection{Interview Content}
Human and AI in-depth interviews were conducted with an identical questionnaire in English (cf. Appendix \ref{text:interview-questions}). The questionnaire concerned questions on politics and democracy (e.g. \textit{Let us talk about democracy. When you think about how democracy works right now in Western countries such as Germany, what are the good things that come to mind?} or \textit{And what do you think “politics” is? How would you define this term?}). Human-led interviews lasted 16 minutes, on average. AI-led interviews lasted 22 minutes, on average.

\subsection{Evaluation Metrics}
We computed a set of quantitative and qualitative measures, designed to evaluate the effectiveness, efficiency, and quality of AI-conducted interviews in comparison to traditional human-conducted interviews. Besides quantitative text-based metrics (\faIcon{search}), we evaluate indicators of participant engagement, response depth, and coherence (\faIcon{user}). Additionally, we gathered survey feedback (\faIcon{file-alt}) on the interview experience from participants in both interview settings.

\paragraph{\faIcon{user} Interviewer behavior: Human coding.} We provided two research assistants with the interviewer guidelines, which outlined desirable and undesirable interviewer behaviors (cf. Appendix \ref{text:coding-guidelines-interviewer}). The research assistants then manually double-coded each conversational turn of the interviewer (e.g., a question) to identify any potential violations of these guidelines. In essence, we assessed whether the human and AI interviewers adhered to the instructions.

\paragraph{\faIcon{user} Interview responses: Human coding.} Two research assistants were provided with a detailed coding manual to assess the quality of the participants' responses (cf. Appendix  \ref{text:coding-guidelines-responses}). They assessed factors such as whether a response directly addressed the question, whether the participant appeared engaged, and the specificity and detail of the response. In essence, we evaluated whether the interviews elicited insightful responses from participants.

\paragraph{\faIcon{search} Interview responses: Computational analysis.}
We computed the Flesch Reading Ease scores on the transcribed interview data to evaluate response readability and length \citep{Flesch1948}. Additionally, we calculated the number of tokens per response to obtain a more granular measure of linguistic complexity and information density.

\paragraph{\faIcon{file-alt} Structured post-interview survey.} After each interview, the respondents were asked to fill out a survey on their experience (cf. Appendix  \ref{text:outcome-questionnaire}).

\paragraph{\faIcon{eye} Real-time problem recording.} During the AI interview, one student from each pair was assigned to observe the other student’s interaction with the AI interviewer. The observer was given a form to document any technical difficulties or other issues the respondent encountered during the interview (cf. Appendix \ref{text:observers-issues}).

\begin{table*}[h!]
\centering
\scalebox{0.9}{
\begin{tabular}{@{}lcccc@{}}
 & $\downarrow{}$ $\uparrow{}$ & \textbf{AI Interviewer} & \textbf{Human Interviewer} & {\footnotesize $\Delta$} \\
\midrule
\\
\faIcon{user} \textbf{Qualitative Assessments} \\
Clarity & $\uparrow{}$ & \textbf{4.3} & 3.9 & {\footnotesize +0.4}\\
Empathy & $\uparrow{}$ & 2.6 & \textbf{2.9} & {\footnotesize -0.3}\\
Engagement & $\uparrow{}$ & 2.6 & \textbf{3.2} & {\footnotesize -0.6}\\
Grammatical correctness & $\uparrow{}$ & \textbf{4.3} & 3.8 & {\footnotesize +0.5} \\
Relevance & $\uparrow{}$ & \textbf{4.6} & 4.3 & {\footnotesize +0.3}\\
Response complexity & $\downarrow{}$ & \textbf{1.9} & 2.1 & {\footnotesize -0.2} \\
Specificity & $\uparrow{}$ & 3.1 & \textbf{3.6} & {\footnotesize -0.5}\\
Tone of answers & $\uparrow{}$ & 3.1 & \textbf{3.3} & {\footnotesize -0.2}\\
\\
\faIcon{search} \textbf{Quantitative Assessments} \\
Tokens per answers & $\uparrow{}$ & \textbf{\wordcountAI} & \wordcountHuman & {\footnotesize +19.58}\\
Readability & $\uparrow{}$ & \textbf{\readbilityaiall} & \readbilityhumanall & {\footnotesize +15.44}\\
\\
\faIcon{file-alt} \textbf{Survey Results} \\
Clarity & $\uparrow{}$ & 1.5 & \textbf{1.9} & {\footnotesize -0.4}\\
Interestingness & $\uparrow{}$ &  2.5 & \textbf{3.9} & {\footnotesize -1.4}\\
Repeatability & $\uparrow{}$ & 2.5 & \textbf{3.6} & {\footnotesize -1.1}\\
Overall Satisfaction & $\uparrow{}$ & \textbf{3.8} & \textbf{3.8} & {\footnotesize +0.0} \\
Understanding & $\uparrow{}$ & 4.0 & \textbf{4.3} & {\footnotesize +0.3}\\
\end{tabular}
}
\caption{Comparison of AI-conducted vs human-conducted interviews: Qualitative assessments \faIcon{user}, quantitative measurements \faIcon{search}, and participant survey \faIcon{file-alt} results where $\Delta$ shows the difference between AI and human scores ($+$ AI performed better and $-$ showing where humans performed better) and we use arrows ($\downarrow{}$ $\uparrow{}$) to indicate the desired direction for each metric - whether a higher ↑ or lower score ↓ is better.}
\label{tab:interview-comparison}
\end{table*}

\section{Findings}

We collected data on six human-led and five AI-conducted interviews. Human-led interviews were audio-recorded and then transcribed. 

Figure~\ref{fig:ai-interface-active-listening} presents an example snippet from an AI conversational interview, showcasing how the interviewer engages in active listening by occasionally repeating the preceding answer, as instructed.

Qualitative inspection of the transcribed data shows that both the AI and human interviewers faithfully followed the provided questionnaire. Manual coding of all interviewer behavior shows that neither humans nor AI always acted in full accordance with the interview guidelines (Figure \ref{fig:errors}). Summarizing across all coded categories, we counted 72 violations per AI interview and 64 violations ($\downarrow{\text{-11.11\%}}$) per human interview, on average. 

While error rates of human and AI interviewers were at similar levels, the nature of the errors differed. Contrary to instructions, human interviewers often failed to engage in active listening, which involves restating the respondent's answer to ensure proper understanding. Specifically, 94 percent of guideline violations related to active listening were committed by human interviewers, compared to only 6 percent by the AI interviewer (cf. Appendix \ref{text:additional-results})). Conversely, and in contrast to internal pre-tests, the AI interviewer predominantly failed to follow the instruction to 'ask follow-up questions when a respondent gives a surprising, unexpected, or unclear answer,' with 88 percent of violations of this rule attributed to the AI interviewer. These findings highlight the challenge of finding the right balance between asking too many and too few follow-up questions in any in-depth interviewing setting. Moreover, the fact that the interviewer model had previously succeeded in asking appropriate follow-up questions during internal tests serves as a reminder that even minor modifications to prompts can lead to unintended side effects.

Another guideline was to avoid any behavior that could bias the respondents' answers. However, despite the instruction to 'not take a position on whether their answers are right or wrong,' the AI interviewer occasionally judged the respondent, typically in an encouraging manner (e.g., 'Your definition of politics is quite insightful', 67 percent attributed to the AI interviewer). In contrast, human interviewers sometimes erred by guiding respondents through associations or suggestions for their answers, accounting for 75 percent of such violations. Overall, while no interviewer setting perfectly adhered to the guidelines, these findings suggest that AI interviewers demonstrate a similar level of effectiveness to human student interviewers in following instructions for in-depth interviewing. However, achieving optimal performance relies on fine-tuning and thoroughly testing model instructions.

Turning from the interviewer's behavior to the participants' responses, we see that both interviewing settings succeeded in eliciting answers from respondents at substantial lengths. In the AI interviewer setting, the average response length was \wordcountAI words. In the human interview setting, the average response length was \wordcountHuman words ($\downarrow{\text{-62.63\%}}$). 

While participants` answers to the AI interviewer were substantial in length, were they also meaningful in substance? The transcribed responses were given to human coders to rate response quality. While we observe minor differences across setting, overall, the ratings indicate a similar response quality. Responses in human and AI interviews were rated as similarly \textit{clear} (i.e., easy to understand), \textit{empathetic} (i.e., sensitive towards the interviewer), \textit{engaged} (i.e., high level of enthusiasm or interest), \textit{complex} (i.e., advanced vocabulary), \textit{grammatically correct }(i.e., error-free), \textit{specific} (i.e., detailed information), and adequate in \textit{tone} (i.e., suitable for the context). 

One particularly important outcome is the assessed relevance of the responses—whether they are useful and directly related to the question asked. Once again, no substantial differences in relevance were observed between AI and human interviews. While these estimates should be interpreted with caution due to the considerable imprecision associated with the small sample size, the findings suggest that engaging with an AI interviewer does not lead to a significant decline in response quality compared to a human interviewer. We interpret this as a proof-of-concept, underscoring the general viability of AI Conversational Interviewing.


Our setup allowed for a close-up investigation of how our AI interviews unfolded in practice. Real-time problem recording during AI interviews showed that respondents interacted seamlessly with our user interface, which resembled familiar chat interfaces, indicating that no learning curve was necessary. Yet, occasionally, the latency of the GPT responses was criticized (e.g. ``\textit{Sometimes the time it takes to produce an answer is unexpectedly long. But it is not really off putting.}'', ``\textit{run time is quite slow, it takes a couple (>5 seconds)}''). While this latency may reflect similar reaction times in human-to-human chat interactions, participants appeared to prefer shorter waiting times when they were aware they were interacting with an AI interviewer.  

\begin{figure*}[!ht]
    \centering
    \includegraphics[width=\textwidth]{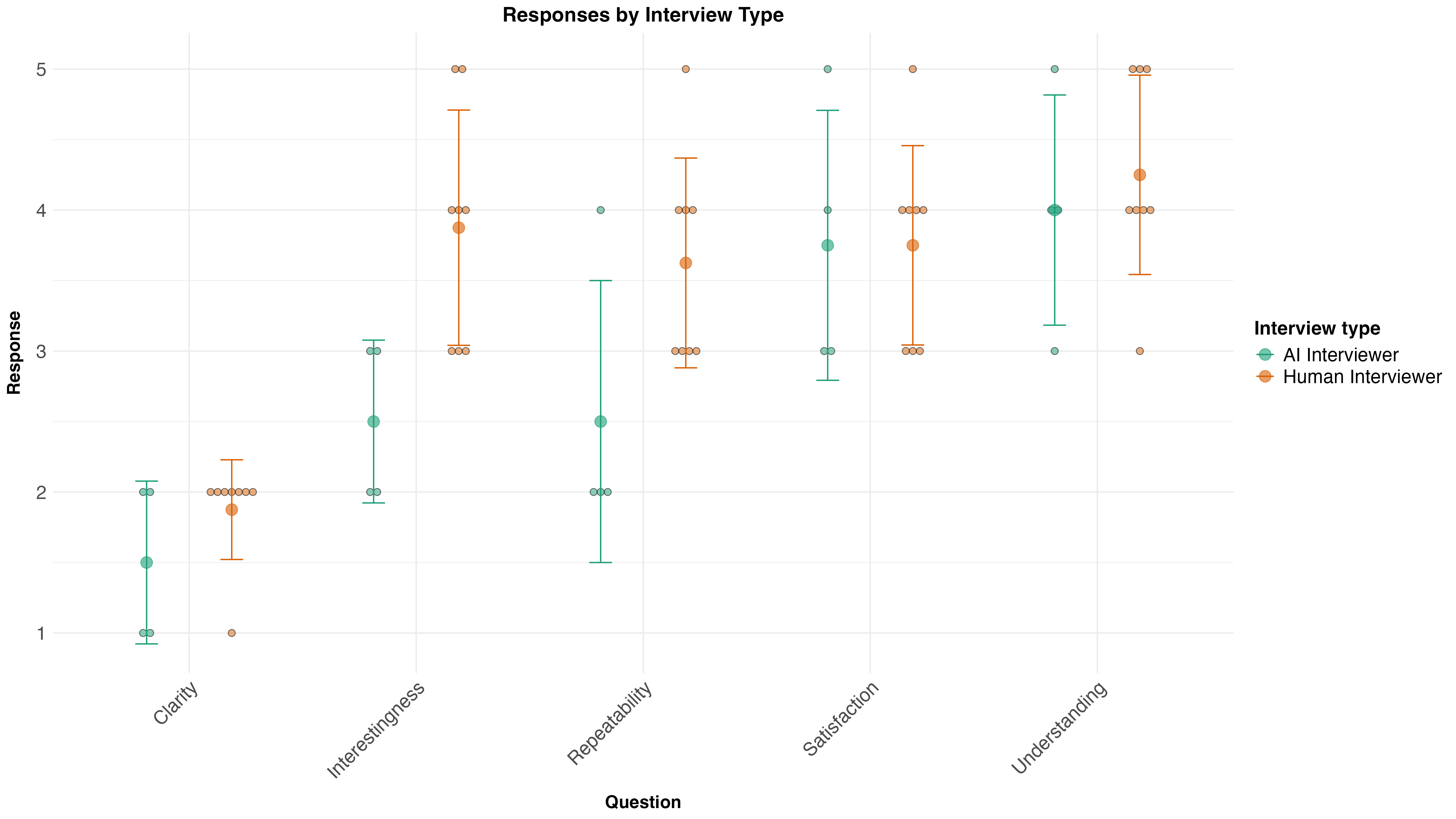}
    \caption{Evaluation for AI (\colorbox[rgb]{0.0,0.8,0.0}{green}) vs Human Interviewers (\colorbox[rgb]{1.0,0.65,0.0}{orange}), showing the scores (y-axis) across different interview assessment criteria for participants' evaluation of interview \faIcon{file-alt} (x-axis).}
    \label{fig:survey_evaluation}
\end{figure*}

Our implementation was voice-assisted, allowing respondents to choose between text and speech for both the interviewer's output and their own input. While no issues were reported with the voice output of the interview questions, the real-time problem recording noted several instances where respondents reported technical issues with audio recording and transcription (``\textit{Some problems with the microphone: Sometimes does not record, speech recognition sometimes recognises words incorrectly}'', ``\textit{small recurring problems with audio recording (not sure if it already runs, accidently stop in recording early}'', ``\textit{recording just stopped completely for a couple seconds and interviewee was kinda mad about it}''). 

Our post-interview survey confirmed these issues. Although five AI interview participants reported trying the audio recording function, only one found it to work sufficiently well to rely on it primarily during the interview. The remaining respondents either partly or primarily preferred to provide written answers to the AI interviewer.

Although unintended, this presents an analytical opportunity to explore differences between written and audio-recorded responses in the AI interviewer setting.  As the survey-methodological literature suggests, the answers of respondents who relied on text input were significantly shorter (on average, \textit{21 tokens} per answer) than the answers by respondents who used audio-recorded throughout the AI interview (\textit{63 tokens} per answer ($\uparrow{\text{+67\%}}$). So, response length markedly varied with input mode. 

However, the survey-methodological literature indicates that audio-recorded responses should not be considered inherently superior but rather qualitatively different from written responses. One student observing a respondent providing written input noted that "\textit{the respondent does not have the opportunity to elaborate in a free way in the written answers. She was very focused on writing good sentences which hindered her in her elaboration}", highlighting the distinct psychological processes associated with each input mode. 

\begin{figure*}[!ht]
    \centering
        \includegraphics[width=\textwidth]{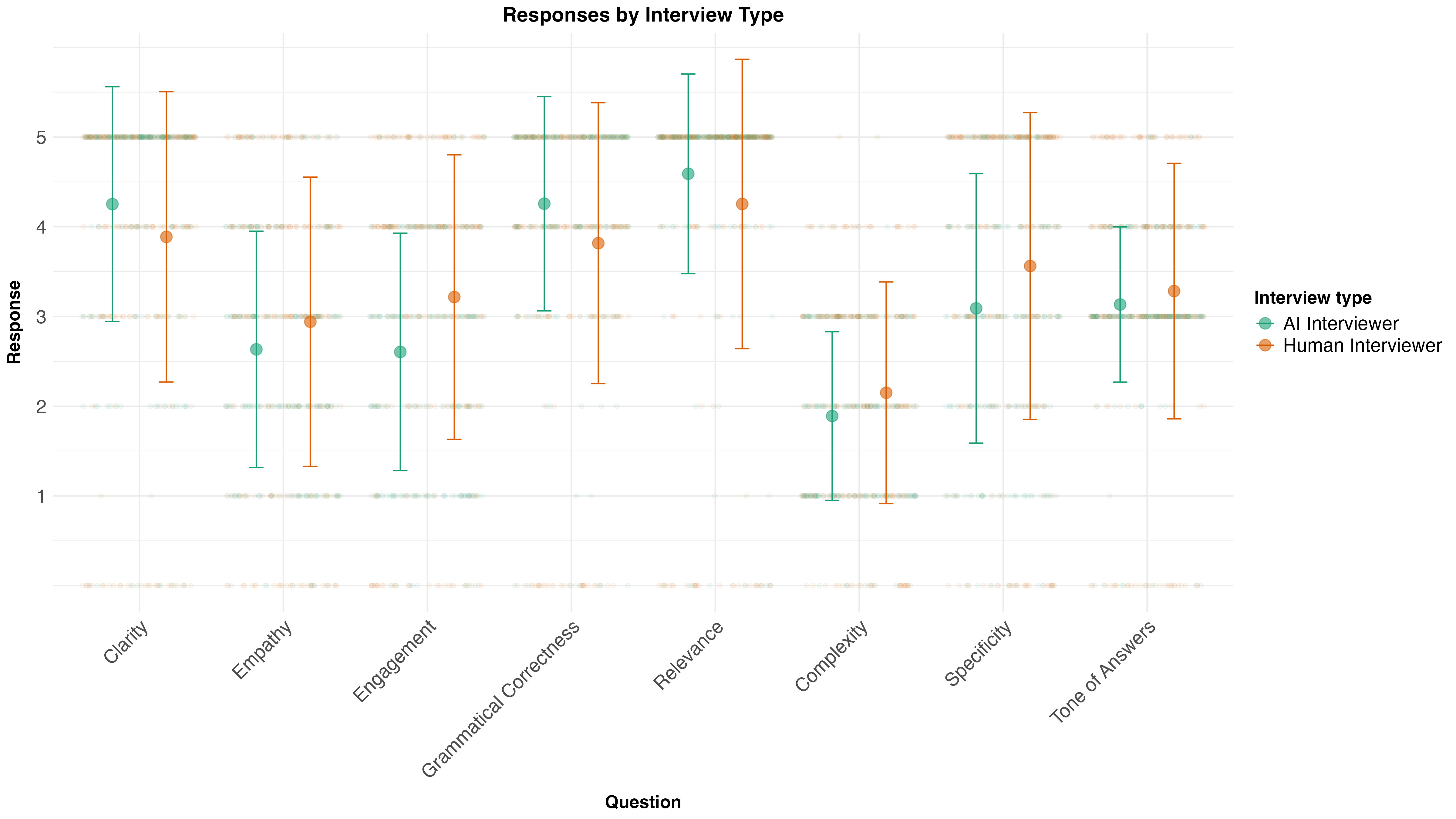}
        \caption{Evaluation for AI (\colorbox[rgb]{0.0,0.8,0.0}{green}) vs Human Interviewers (\colorbox[rgb]{1.0,0.65,0.0}{orange}), showing the scores (y-axis) across different interview assessment criteria for human-rated response quality \faIcon{user} (x-axis).}
        \label{fig:figure1}
\end{figure*}

Further qualitative observation indeed suggests that text-based inputs encourage respondents to think before writing, whereas audio recording tends to prompt respondents to ``think out loud'', allowing them to develop their thoughts while speaking (see Appendix \ref{text:observers-issues} for an example). 
The response styles associated with audio- and text-based input modes are also reflected in objective measures we extracted from the transcribed interview data. Text-based AI interviews achieved a Flesch Reading Ease score of \readbilityaiall while the fully audio-based AI interview scored at \readbilityaiwvoice ($\uparrow{\text{+62,22\%}}$) (Flesch Reading Ease score for human interviews: 62; higher values indicate higher readability). Hence, compared to text input, audio input in AI interviews may be associated with longer but less elaborate answers. 
How did respondents experience the interviews? Participants felt that both the human and AI interviewers were clear in their questions and that each understood their responses (Figure \ref{fig:survey_evaluation}). Respondents in both settings left the interview satisfied. However, participants found the AI interview less interesting and were less likely to repeat it, possibly due to the technical problems with the audio recording. While emphasizing that a satisfactory interview hinges on a flawless technical implementation of the interview process, these findings suggest that the absence of a human interviewer does not necessarily need to go along with a deteriorated interview experience for the respondents. 

\section{Discussion and Recommendations}

Applying the questionnaire to a student sample with both human and AI interviewers demonstrates the general viability of AI Conversational Interviewing. When properly implemented, AI Conversational Interviewing can collect high-quality data. A comprehensive set of qualitative and quantitative metrics suggests that AI interviewing maintains quality comparable to that of human interviewing, but at significantly lower costs, thereby making in-depth interviews more scalable.

Although these findings highlight the potential of AI Conversational Interviewing, the success of the method depends on its precise implementation. Based on our comprehensive analysis, we present five recommendations for the future development and employment of AI-driven in-depth interviews:

\paragraph{Leverage existing knowledge.} When specifying desired interviewer behavior, it is crucial to draw on established principles from survey methodology. These practices, developed through extensive research and practical experience, offer proven guidelines for effective implementation.

\paragraph{Context-specific definition of desired interviewer behavior.} It is crucial to make deliberate judgment calls to tailor the desired interviewer behavior to your specific research context. This may involve decisions on aspects such as the importance or frequency of follow-up questions, the depth of probing on certain topics, or the level of formality in the interview tone (for example, \citet{weidmann_dialing_2024} demonstrated the effectiveness of empathy prompting). Each research project may require a unique approach to AI interviewer behavior to ensure the collection of appropriate data.

\paragraph{Consider user experience.} The interface through which participants interact with the AI interviewer is crucial to the success of the interview. It is essential to rely on familiar and intuitive user interfaces that minimize cognitive load and technical barriers. Well-designed interfaces enable participants to focus on providing thoughtful responses rather than being distracted by technical difficulties.

\paragraph{Careful prompting.} The prompts provided to the AI interviewer are crucial to its performance. Conduct thorough pre-testing to ensure that the AI's behavior aligns with your established guidelines. It is important to consider the potential unintended side-effects of modifying prompts, as even minor adjustments can lead to significant changes in interviewer behavior or question interpretation \citep{tam2024letspeakfreelystudy, sclar2024quantifying, zhu2024promptrobustevaluatingrobustnesslarge}.

\paragraph{Input mode matters.} Recognize that the chosen input mode (e.g., text or speech) will significantly influence participant behavior by eliciting different psychological responses. Response patterns may vary across several outcomes, sometimes in contrasting ways. For instance, spoken responses might be longer but less detailed, while written responses may be shorter yet more concise and thoughtfully constructed. The choice of input mode should be made with careful consideration of your research objectives and the type of data you aim to collect.

\section{Conclusion}
Our research contributes to the growing field of AI-supported interviewing by offering initial insights through an in-depth evaluation process. We assessed AI performance using a variety of quantitative and qualitative evaluation methods, documenting the challenges participants faced and comparing AI-conducted interviews with human-led ones. To ensure transparency, we have made our pipeline, questions, and data publicly available. Based on our preliminary findings, we propose five areas for consideration in future implementations: integrating established survey methodology principles, adapting AI behavior to different contexts, designing user-friendly interfaces, conducting comprehensive pre-testing, and being aware of input mode effects. While our results highlight the potential of AI Conversational Interviewing, it is important to recognize that outcomes are heavily dependent on the specific implementation methods used.

\section*{Limitations}
Several limitations reflect our study's design of a close-up monitoring of AI interviewing in practice. The study's small sample limits the generalizability of the findings. Our decision to have students monitor the AI interviewing process impedes investigating whether the absence of a human being fosters respondents' proclivity to discuss sensitive topics which may be an additional advantage of AI Conversational Interviewing. Our participants were students with an interest in survey methodology which may have been more motivated than ordinary participants. Furthermore, the use of a closed model restricts the study's replicability compared to the transparency that could be achieved with an open-source model \citep{spirling2023open}. We chose GPT-4 because it was the state of the art at the time of the interviews and offered social science researchers the most accessible opportunity for application \citep{palmer2024using}. By showing the pitfalls of the best-performing model across several benchmarks, we aimed to provide a starting point for an open discussion on this type of model. For future research, we plan to compare the capabilities of different models, including strong open-source models such as Llama 3.1 \citep{dubey2024llama3herdmodels}, to provide a more comprehensive and application-oriented view of AI interviewing techniques. Finally, our study concerned  collecting data via AI Conversational interviews and not its analysis where researchers may rely on computational methods for text analysis \citep{baden2022three, banks2018review, dimaggio2015adapting, grimmer_text_2022}. 

\section*{Ethics Statement}

We affirm that our research adheres to the \href{https://www.aclweb.org/portal/content/acl-code-ethics}{ACL Ethics Policy}. To protect participant privacy, we ensure that no individuals are identifiable. To maximize the public value of our work, we make all underlying data and source code openly available for reuse. We declare that no conflicts of interest could influence the study’s outcomes, interpretations, or conclusions. All funding sources supporting this research are acknowledged in the acknowledgments section. Furthermore, we have rigorously documented our methodology, experiments, and results to enhance the replicability of our findings. 

\section*{Acknowledgements}

We thank Laura Kiemes and Valeriya Barakhvostova for excellent research assistance. We are grateful for helpful comments we received at I2SC Saarbrücken Kick-Off Event, the Mainz Workshop on Citizen Perspectives on Demcracy, and the LMU MCMP seminar series. Alexander Wuttke was funded by LMU's Young Researcher Support Fund. Matthias Aßenmacher was funded by the Deutsche Forschungsgemeinschaft (DFG, German Research Foundation) under the National Research Data Infrastructure – NFDI 27/1 - 460037581.

\bibliography{acl_latex}

\begin{thebibliography}{45}
\providecommand{\natexlab}[1]{#1}

\bibitem[{Adams(2015)}]{newcomer_conducting_2015}
William~C. Adams. 2015.
\newblock \href {https://doi.org/10.1002/9781119171386.ch19} {Conducting {Semi}‐{Structured} {Interviews}}.
\newblock In Kathryn~E. Newcomer, Harry~P. Hatry, and Joseph~S. Wholey, editors, \emph{Handbook of {Practical} {Program} {Evaluation}}, 1 edition, pages 492--505. Wiley.

\bibitem[{Adeoye-Olatunde and Olenik(2021)}]{adeoye2021research}
Omolola~A Adeoye-Olatunde and Nicole~L Olenik. 2021.
\newblock Research and scholarly methods: Semi-structured interviews.
\newblock \emph{Journal of the american college of clinical pharmacy}, 4(10):1358--1367.

\bibitem[{Atkeson et~al.(2014)Atkeson, Adams, and Alvarez}]{atkeson_nonresponse_2014}
Lonna~Rae Atkeson, Alex~N. Adams, and R.~Michael Alvarez. 2014.
\newblock \href {https://doi.org/10.1093/pan/mpt049} {Nonresponse and {Mode} {Effects} in {Self}- and {Interviewer}-{Administered} {Surveys}}.
\newblock \emph{Political Analysis}, 22(3):304--320.

\bibitem[{Baburajan et~al.(2022)Baburajan, e~Silva, and Pereira}]{baburajan2022open}
Vishnu Baburajan, Jo{\~a}o de~Abreu e~Silva, and Francisco~Camara Pereira. 2022.
\newblock Open vs closed-ended questions in attitudinal surveys--comparing, combining, and interpreting using natural language processing.
\newblock \emph{Transportation research part C: emerging technologies}, 137:103589.

\bibitem[{Baden et~al.(2022)Baden, Pipal, Schoonvelde, and van~der Velden}]{baden2022three}
Christian Baden, Christian Pipal, Martijn Schoonvelde, and Mariken AC~G van~der Velden. 2022.
\newblock Three gaps in computational text analysis methods for social sciences: A research agenda.
\newblock \emph{Communication Methods and Measures}, 16(1):1--18.

\bibitem[{Banks et~al.(2018)Banks, Woznyj, Wesslen, and Ross}]{banks2018review}
George~C Banks, Haley~M Woznyj, Ryan~S Wesslen, and Roxanne~L Ross. 2018.
\newblock A review of best practice recommendations for text analysis in r (and a user-friendly app).
\newblock \emph{Journal of Business and Psychology}, 33:445--459.

\bibitem[{Cai et~al.(2024)Cai, Duan, Haslett, Wang, and Pickering}]{cai2024largelanguagemodelsresemble}
Zhenguang~G. Cai, Xufeng Duan, David~A. Haslett, Shuqi Wang, and Martin~J. Pickering. 2024.
\newblock \href {https://arxiv.org/abs/2303.08014} {Do large language models resemble humans in language use?}
\newblock \emph{Preprint}, arXiv:2303.08014.

\bibitem[{Chang et~al.(2021)Chang, Ferguson, Rothschild, and Page}]{chang2021ambivalence}
Arturo Chang, Thomas Ferguson, Jacob Rothschild, and Benjamin Page. 2021.
\newblock Ambivalence about international trade in open-and closed-ended survey responses.
\newblock \emph{Institute for New Economic Thinking Working Paper Series}, 162.

\bibitem[{Chopra and Haaland(2023)}]{chopra_conducting_2023}
Felix Chopra and Ingar Haaland. 2023.
\newblock \href {https://papers.ssrn.com/abstract=4572954} {Conducting {Qualitative} {Interviews} with {AI}}.

\bibitem[{Costello et~al.(2024)Costello, Pennycook, and Rand}]{costello2024durably}
Thomas~H Costello, Gordon Pennycook, and David~G Rand. 2024.
\newblock Durably reducing conspiracy beliefs through dialogues with ai.
\newblock \emph{Science}, 385(6714):eadq1814.

\bibitem[{Davis et~al.(2024)Davis, Conrad, Dong, Mesa, Lee, and Johnson}]{davis2024ounce}
Rachel~E Davis, Frederick~G Conrad, Shaohua Dong, Anna Mesa, Sunghee Lee, and Timothy~P Johnson. 2024.
\newblock An ounce of prevention: using conversational interviewing and avoiding agreement response scales to prevent acquiescence.
\newblock \emph{Quality \& Quantity}, 58(1):471--495.

\bibitem[{di~San~Pietro et~al.(2023)di~San~Pietro, Frau, Mangiaterra, and Bambini}]{di2023pragmatic}
Chiara~Barattieri di~San~Pietro, Federico Frau, Veronica Mangiaterra, and Valentina Bambini. 2023.
\newblock The pragmatic profile of chatgpt: assessing the pragmatic skills of a conversational agent.
\newblock \emph{PsyArXiv}.

\bibitem[{DiMaggio(2015)}]{dimaggio2015adapting}
Paul DiMaggio. 2015.
\newblock Adapting computational text analysis to social science (and vice versa).
\newblock \emph{Big Data \& Society}, 2(2):2053951715602908.

\bibitem[{Dubey et~al.(2024)Dubey, Jauhri, Pandey, Kadian, Al-Dahle, Letman, Mathur, Schelten, Yang, Fan, Goyal, Hartshorn, Yang, Mitra, Sravankumar, Korenev, Hinsvark, Rao, Zhang, Rodriguez, Gregerson, Spataru, Roziere, Biron, Tang, Chern, Caucheteux, Nayak, Bi, Marra, McConnell, Keller, Touret, Wu, Wong, Ferrer, Nikolaidis, Allonsius, Song, Pintz, Livshits, Esiobu, Choudhary, Mahajan, Garcia-Olano, Perino, Hupkes, Lakomkin, AlBadawy, Lobanova, Dinan, Smith, Radenovic, Zhang, Synnaeve, Lee, Anderson, Nail, Mialon, Pang, Cucurell, Nguyen, Korevaar, Xu, Touvron, Zarov, Ibarra, Kloumann, Misra, Evtimov, Copet, Lee, Geffert, Vranes, Park, Mahadeokar, Shah, van~der Linde, Billock, Hong, Lee, Fu, Chi, Huang, Liu, Wang, Yu, Bitton, Spisak, Park, Rocca, Johnstun, Saxe, Jia, Alwala, Upasani, Plawiak, Li, Heafield, Stone, El-Arini, Iyer, Malik, Chiu, Bhalla, Rantala-Yeary, van~der Maaten, Chen, Tan, Jenkins, Martin, Madaan, Malo, Blecher, Landzaat, de~Oliveira, Muzzi, Pasupuleti, Singh, Paluri, Kardas, Oldham, Rita,
  Pavlova, Kambadur, Lewis, Si, Singh, Hassan, Goyal, Torabi, Bashlykov, Bogoychev, Chatterji, Duchenne, Çelebi, Alrassy, Zhang, Li, Vasic, Weng, Bhargava, Dubal, Krishnan, Koura, Xu, He, Dong, Srinivasan, Ganapathy, Calderer, Cabral, Stojnic, Raileanu, Girdhar, Patel, Sauvestre, Polidoro, Sumbaly, Taylor, Silva, Hou, Wang, Hosseini, Chennabasappa, Singh, Bell, Kim, Edunov, Nie, Narang, Raparthy, Shen, Wan, Bhosale, Zhang, Vandenhende, Batra, Whitman, Sootla, Collot, Gururangan, Borodinsky, Herman, Fowler, Sheasha, Georgiou, Scialom, Speckbacher, Mihaylov, Xiao, Karn, Goswami, Gupta, Ramanathan, Kerkez, Gonguet, Do, Vogeti, Petrovic, Chu, Xiong, Fu, Meers, Martinet, Wang, Tan, Xie, Jia, Wang, Goldschlag, Gaur, Babaei, Wen, Song, Zhang, Li, Mao, Coudert, Yan, Chen, Papakipos, Singh, Grattafiori, Jain, Kelsey, Shajnfeld, Gangidi, Victoria, Goldstand, Menon, Sharma, Boesenberg, Vaughan, Baevski, Feinstein, Kallet, Sangani, Yunus, Lupu, Alvarado, Caples, Gu, Ho, Poulton, Ryan, Ramchandani, Franco, Saraf,
  Chowdhury, Gabriel, Bharambe, Eisenman, Yazdan, James, Maurer, Leonhardi, Huang, Loyd, Paola, Paranjape, Liu, Wu, Ni, Hancock, Wasti, Spence, Stojkovic, Gamido, Montalvo, Parker, Burton, Mejia, Wang, Kim, Zhou, Hu, Chu, Cai, Tindal, Feichtenhofer, Civin, Beaty, Kreymer, Li, Wyatt, Adkins, Xu, Testuggine, David, Parikh, Liskovich, Foss, Wang, Le, Holland, Dowling, Jamil, Montgomery, Presani, Hahn, Wood, Brinkman, Arcaute, Dunbar, Smothers, Sun, Kreuk, Tian, Ozgenel, Caggioni, Guzmán, Kanayet, Seide, Florez, Schwarz, Badeer, Swee, Halpern, Thattai, Herman, Sizov, Guangyi, Zhang, Lakshminarayanan, Shojanazeri, Zou, Wang, Zha, Habeeb, Rudolph, Suk, Aspegren, Goldman, Damlaj, Molybog, Tufanov, Veliche, Gat, Weissman, Geboski, Kohli, Asher, Gaya, Marcus, Tang, Chan, Zhen, Reizenstein, Teboul, Zhong, Jin, Yang, Cummings, Carvill, Shepard, McPhie, Torres, Ginsburg, Wang, Wu, U, Saxena, Prasad, Khandelwal, Zand, Matosich, Veeraraghavan, Michelena, Li, Huang, Chawla, Lakhotia, Huang, Chen, Garg, A, Silva, Bell,
  Zhang, Guo, Yu, Moshkovich, Wehrstedt, Khabsa, Avalani, Bhatt, Tsimpoukelli, Mankus, Hasson, Lennie, Reso, Groshev, Naumov, Lathi, Keneally, Seltzer, Valko, Restrepo, Patel, Vyatskov, Samvelyan, Clark, Macey, Wang, Hermoso, Metanat, Rastegari, Bansal, Santhanam, Parks, White, Bawa, Singhal, Egebo, Usunier, Laptev, Dong, Zhang, Cheng, Chernoguz, Hart, Salpekar, Kalinli, Kent, Parekh, Saab, Balaji, Rittner, Bontrager, Roux, Dollar, Zvyagina, Ratanchandani, Yuvraj, Liang, Alao, Rodriguez, Ayub, Murthy, Nayani, Mitra, Li, Hogan, Battey, Wang, Maheswari, Howes, Rinott, Bondu, Datta, Chugh, Hunt, Dhillon, Sidorov, Pan, Verma, Yamamoto, Ramaswamy, Lindsay, Lindsay, Feng, Lin, Zha, Shankar, Zhang, Zhang, Wang, Agarwal, Sajuyigbe, Chintala, Max, Chen, Kehoe, Satterfield, Govindaprasad, Gupta, Cho, Virk, Subramanian, Choudhury, Goldman, Remez, Glaser, Best, Kohler, Robinson, Li, Zhang, Matthews, Chou, Shaked, Vontimitta, Ajayi, Montanez, Mohan, Kumar, Mangla, Albiero, Ionescu, Poenaru, Mihailescu, Ivanov, Li, Wang,
  Jiang, Bouaziz, Constable, Tang, Wang, Wu, Wang, Xia, Wu, Gao, Chen, Hu, Jia, Qi, Li, Zhang, Zhang, Adi, Nam, Yu, Wang, Hao, Qian, He, Rait, DeVito, Rosnbrick, Wen, Yang, and Zhao}]{dubey2024llama3herdmodels}
Abhimanyu Dubey, Abhinav Jauhri, Abhinav Pandey, Abhishek Kadian, Ahmad Al-Dahle, Aiesha Letman, Akhil Mathur, Alan Schelten, Amy Yang, Angela Fan, Anirudh Goyal, Anthony Hartshorn, Aobo Yang, Archi Mitra, Archie Sravankumar, Artem Korenev, Arthur Hinsvark, Arun Rao, Aston Zhang, Aurelien Rodriguez, Austen Gregerson, Ava Spataru, Baptiste Roziere, Bethany Biron, Binh Tang, Bobbie Chern, Charlotte Caucheteux, Chaya Nayak, Chloe Bi, Chris Marra, Chris McConnell, Christian Keller, Christophe Touret, Chunyang Wu, Corinne Wong, Cristian~Canton Ferrer, Cyrus Nikolaidis, Damien Allonsius, Daniel Song, Danielle Pintz, Danny Livshits, David Esiobu, Dhruv Choudhary, Dhruv Mahajan, Diego Garcia-Olano, Diego Perino, Dieuwke Hupkes, Egor Lakomkin, Ehab AlBadawy, Elina Lobanova, Emily Dinan, Eric~Michael Smith, Filip Radenovic, Frank Zhang, Gabriel Synnaeve, Gabrielle Lee, Georgia~Lewis Anderson, Graeme Nail, Gregoire Mialon, Guan Pang, Guillem Cucurell, Hailey Nguyen, Hannah Korevaar, Hu~Xu, Hugo Touvron, Iliyan Zarov,
  Imanol~Arrieta Ibarra, Isabel Kloumann, Ishan Misra, Ivan Evtimov, Jade Copet, Jaewon Lee, Jan Geffert, Jana Vranes, Jason Park, Jay Mahadeokar, Jeet Shah, Jelmer van~der Linde, Jennifer Billock, Jenny Hong, Jenya Lee, Jeremy Fu, Jianfeng Chi, Jianyu Huang, Jiawen Liu, Jie Wang, Jiecao Yu, Joanna Bitton, Joe Spisak, Jongsoo Park, Joseph Rocca, Joshua Johnstun, Joshua Saxe, Junteng Jia, Kalyan~Vasuden Alwala, Kartikeya Upasani, Kate Plawiak, Ke~Li, Kenneth Heafield, Kevin Stone, Khalid El-Arini, Krithika Iyer, Kshitiz Malik, Kuenley Chiu, Kunal Bhalla, Lauren Rantala-Yeary, Laurens van~der Maaten, Lawrence Chen, Liang Tan, Liz Jenkins, Louis Martin, Lovish Madaan, Lubo Malo, Lukas Blecher, Lukas Landzaat, Luke de~Oliveira, Madeline Muzzi, Mahesh Pasupuleti, Mannat Singh, Manohar Paluri, Marcin Kardas, Mathew Oldham, Mathieu Rita, Maya Pavlova, Melanie Kambadur, Mike Lewis, Min Si, Mitesh~Kumar Singh, Mona Hassan, Naman Goyal, Narjes Torabi, Nikolay Bashlykov, Nikolay Bogoychev, Niladri Chatterji, Olivier
  Duchenne, Onur Çelebi, Patrick Alrassy, Pengchuan Zhang, Pengwei Li, Petar Vasic, Peter Weng, Prajjwal Bhargava, Pratik Dubal, Praveen Krishnan, Punit~Singh Koura, Puxin Xu, Qing He, Qingxiao Dong, Ragavan Srinivasan, Raj Ganapathy, Ramon Calderer, Ricardo~Silveira Cabral, Robert Stojnic, Roberta Raileanu, Rohit Girdhar, Rohit Patel, Romain Sauvestre, Ronnie Polidoro, Roshan Sumbaly, Ross Taylor, Ruan Silva, Rui Hou, Rui Wang, Saghar Hosseini, Sahana Chennabasappa, Sanjay Singh, Sean Bell, Seohyun~Sonia Kim, Sergey Edunov, Shaoliang Nie, Sharan Narang, Sharath Raparthy, Sheng Shen, Shengye Wan, Shruti Bhosale, Shun Zhang, Simon Vandenhende, Soumya Batra, Spencer Whitman, Sten Sootla, Stephane Collot, Suchin Gururangan, Sydney Borodinsky, Tamar Herman, Tara Fowler, Tarek Sheasha, Thomas Georgiou, Thomas Scialom, Tobias Speckbacher, Todor Mihaylov, Tong Xiao, Ujjwal Karn, Vedanuj Goswami, Vibhor Gupta, Vignesh Ramanathan, Viktor Kerkez, Vincent Gonguet, Virginie Do, Vish Vogeti, Vladan Petrovic, Weiwei Chu,
  Wenhan Xiong, Wenyin Fu, Whitney Meers, Xavier Martinet, Xiaodong Wang, Xiaoqing~Ellen Tan, Xinfeng Xie, Xuchao Jia, Xuewei Wang, Yaelle Goldschlag, Yashesh Gaur, Yasmine Babaei, Yi~Wen, Yiwen Song, Yuchen Zhang, Yue Li, Yuning Mao, Zacharie~Delpierre Coudert, Zheng Yan, Zhengxing Chen, Zoe Papakipos, Aaditya Singh, Aaron Grattafiori, Abha Jain, Adam Kelsey, Adam Shajnfeld, Adithya Gangidi, Adolfo Victoria, Ahuva Goldstand, Ajay Menon, Ajay Sharma, Alex Boesenberg, Alex Vaughan, Alexei Baevski, Allie Feinstein, Amanda Kallet, Amit Sangani, Anam Yunus, Andrei Lupu, Andres Alvarado, Andrew Caples, Andrew Gu, Andrew Ho, Andrew Poulton, Andrew Ryan, Ankit Ramchandani, Annie Franco, Aparajita Saraf, Arkabandhu Chowdhury, Ashley Gabriel, Ashwin Bharambe, Assaf Eisenman, Azadeh Yazdan, Beau James, Ben Maurer, Benjamin Leonhardi, Bernie Huang, Beth Loyd, Beto~De Paola, Bhargavi Paranjape, Bing Liu, Bo~Wu, Boyu Ni, Braden Hancock, Bram Wasti, Brandon Spence, Brani Stojkovic, Brian Gamido, Britt Montalvo, Carl
  Parker, Carly Burton, Catalina Mejia, Changhan Wang, Changkyu Kim, Chao Zhou, Chester Hu, Ching-Hsiang Chu, Chris Cai, Chris Tindal, Christoph Feichtenhofer, Damon Civin, Dana Beaty, Daniel Kreymer, Daniel Li, Danny Wyatt, David Adkins, David Xu, Davide Testuggine, Delia David, Devi Parikh, Diana Liskovich, Didem Foss, Dingkang Wang, Duc Le, Dustin Holland, Edward Dowling, Eissa Jamil, Elaine Montgomery, Eleonora Presani, Emily Hahn, Emily Wood, Erik Brinkman, Esteban Arcaute, Evan Dunbar, Evan Smothers, Fei Sun, Felix Kreuk, Feng Tian, Firat Ozgenel, Francesco Caggioni, Francisco Guzmán, Frank Kanayet, Frank Seide, Gabriela~Medina Florez, Gabriella Schwarz, Gada Badeer, Georgia Swee, Gil Halpern, Govind Thattai, Grant Herman, Grigory Sizov, Guangyi, Zhang, Guna Lakshminarayanan, Hamid Shojanazeri, Han Zou, Hannah Wang, Hanwen Zha, Haroun Habeeb, Harrison Rudolph, Helen Suk, Henry Aspegren, Hunter Goldman, Ibrahim Damlaj, Igor Molybog, Igor Tufanov, Irina-Elena Veliche, Itai Gat, Jake Weissman, James
  Geboski, James Kohli, Japhet Asher, Jean-Baptiste Gaya, Jeff Marcus, Jeff Tang, Jennifer Chan, Jenny Zhen, Jeremy Reizenstein, Jeremy Teboul, Jessica Zhong, Jian Jin, Jingyi Yang, Joe Cummings, Jon Carvill, Jon Shepard, Jonathan McPhie, Jonathan Torres, Josh Ginsburg, Junjie Wang, Kai Wu, Kam~Hou U, Karan Saxena, Karthik Prasad, Kartikay Khandelwal, Katayoun Zand, Kathy Matosich, Kaushik Veeraraghavan, Kelly Michelena, Keqian Li, Kun Huang, Kunal Chawla, Kushal Lakhotia, Kyle Huang, Lailin Chen, Lakshya Garg, Lavender A, Leandro Silva, Lee Bell, Lei Zhang, Liangpeng Guo, Licheng Yu, Liron Moshkovich, Luca Wehrstedt, Madian Khabsa, Manav Avalani, Manish Bhatt, Maria Tsimpoukelli, Martynas Mankus, Matan Hasson, Matthew Lennie, Matthias Reso, Maxim Groshev, Maxim Naumov, Maya Lathi, Meghan Keneally, Michael~L. Seltzer, Michal Valko, Michelle Restrepo, Mihir Patel, Mik Vyatskov, Mikayel Samvelyan, Mike Clark, Mike Macey, Mike Wang, Miquel~Jubert Hermoso, Mo~Metanat, Mohammad Rastegari, Munish Bansal, Nandhini
  Santhanam, Natascha Parks, Natasha White, Navyata Bawa, Nayan Singhal, Nick Egebo, Nicolas Usunier, Nikolay~Pavlovich Laptev, Ning Dong, Ning Zhang, Norman Cheng, Oleg Chernoguz, Olivia Hart, Omkar Salpekar, Ozlem Kalinli, Parkin Kent, Parth Parekh, Paul Saab, Pavan Balaji, Pedro Rittner, Philip Bontrager, Pierre Roux, Piotr Dollar, Polina Zvyagina, Prashant Ratanchandani, Pritish Yuvraj, Qian Liang, Rachad Alao, Rachel Rodriguez, Rafi Ayub, Raghotham Murthy, Raghu Nayani, Rahul Mitra, Raymond Li, Rebekkah Hogan, Robin Battey, Rocky Wang, Rohan Maheswari, Russ Howes, Ruty Rinott, Sai~Jayesh Bondu, Samyak Datta, Sara Chugh, Sara Hunt, Sargun Dhillon, Sasha Sidorov, Satadru Pan, Saurabh Verma, Seiji Yamamoto, Sharadh Ramaswamy, Shaun Lindsay, Shaun Lindsay, Sheng Feng, Shenghao Lin, Shengxin~Cindy Zha, Shiva Shankar, Shuqiang Zhang, Shuqiang Zhang, Sinong Wang, Sneha Agarwal, Soji Sajuyigbe, Soumith Chintala, Stephanie Max, Stephen Chen, Steve Kehoe, Steve Satterfield, Sudarshan Govindaprasad, Sumit Gupta,
  Sungmin Cho, Sunny Virk, Suraj Subramanian, Sy~Choudhury, Sydney Goldman, Tal Remez, Tamar Glaser, Tamara Best, Thilo Kohler, Thomas Robinson, Tianhe Li, Tianjun Zhang, Tim Matthews, Timothy Chou, Tzook Shaked, Varun Vontimitta, Victoria Ajayi, Victoria Montanez, Vijai Mohan, Vinay~Satish Kumar, Vishal Mangla, Vítor Albiero, Vlad Ionescu, Vlad Poenaru, Vlad~Tiberiu Mihailescu, Vladimir Ivanov, Wei Li, Wenchen Wang, Wenwen Jiang, Wes Bouaziz, Will Constable, Xiaocheng Tang, Xiaofang Wang, Xiaojian Wu, Xiaolan Wang, Xide Xia, Xilun Wu, Xinbo Gao, Yanjun Chen, Ye~Hu, Ye~Jia, Ye~Qi, Yenda Li, Yilin Zhang, Ying Zhang, Yossi Adi, Youngjin Nam, Yu, Wang, Yuchen Hao, Yundi Qian, Yuzi He, Zach Rait, Zachary DeVito, Zef Rosnbrick, Zhaoduo Wen, Zhenyu Yang, and Zhiwei Zhao. 2024.
\newblock \href {https://arxiv.org/abs/2407.21783} {The llama 3 herd of models}.
\newblock \emph{Preprint}, arXiv:2407.21783.

\bibitem[{Duck-Mayr and Montgomery(2023)}]{duck2023ends}
JBrandon Duck-Mayr and Jacob Montgomery. 2023.
\newblock Ends against the middle: Measuring latent traits when opposites respond the same way for antithetical reasons.
\newblock \emph{Political Analysis}, 31(4):606--625.

\bibitem[{Esses and Maio(2002)}]{esses2002expanding}
Victoria~M Esses and Gregory~R Maio. 2002.
\newblock Expanding the assessment of attitude components and structure: The benefits of open-ended measures.
\newblock \emph{European review of social psychology}, 12(1):71--101.

\bibitem[{Flesch(1948)}]{Flesch1948}
Rudolph Flesch. 1948.
\newblock \href {http://libezproxy.open.ac.uk/login?url=http://search.ebscohost.com.libezproxy.open.ac.uk/login.aspx?direct=true&db=pdh&AN=apl-32-3-221&site=ehost-live&scope=site} {A new readability yardstick.}
\newblock \emph{Journal of Applied Psychology}, 32(3):p221 -- 233.

\bibitem[{Gavras et~al.(2022)Gavras, Höhne, Blom, and Schoen}]{gavras_innovating_2022}
Konstantin Gavras, Jan~Karem Höhne, Annelies~G. Blom, and Harald Schoen. 2022.
\newblock \href {https://doi.org/10.1111/rssa.12807} {Innovating the {Collection} of {Open}-{Ended} {Answers}: {The} {Linguistic} and {Content} {Characteristics} of {Written} and {Oral} {Answers} to {Political} {Attitude} {Questions}}.
\newblock \emph{Journal of the Royal Statistical Society Series A: Statistics in Society}, 185(3):872--890.

\bibitem[{Grimmer et~al.(2022)Grimmer, Roberts, and Stewart}]{grimmer_text_2022}
Justin Grimmer, Margaret~E Roberts, and Brandon~M Stewart. 2022.
\newblock \href {https://press.princeton.edu/books/hardcover/9780691207544/text-as-data} {\emph{Text as data: A new framework for machine learning and the social sciences}}.
\newblock Princeton University Press.

\bibitem[{Groves(2009)}]{groves_survey_2009}
Robert~M. Groves, editor. 2009.
\newblock \emph{Survey methodology}, 2nd ed edition.
\newblock Wiley series in survey methodology. Wiley, Hoboken, N.J.
\newblock OCLC: ocn302189175.

\bibitem[{Höhne et~al.(2024)Höhne, Kern, Gavras, and Schlosser}]{hohne_sound_2024}
Jan~Karem Höhne, Christoph Kern, Konstantin Gavras, and Stephan Schlosser. 2024.
\newblock \href {https://doi.org/10.1007/s11135-023-01776-8} {The sound of respondents: predicting respondents’ level of interest in questions with voice data in smartphone surveys}.
\newblock \emph{Quality \& Quantity}, 58(3):2907--2927.

\bibitem[{Jeong et~al.(2023)Jeong, Aggarwal, Robinson, Kumar, Spearot, and Park}]{jeong2023exhaustive}
Dahyeon Jeong, Shilpa Aggarwal, Jonathan Robinson, Naresh Kumar, Alan Spearot, and David~Sungho Park. 2023.
\newblock Exhaustive or exhausting? evidence on respondent fatigue in long surveys.
\newblock \emph{Journal of Development Economics}, 161:102992.

\bibitem[{Kallio et~al.(2016)Kallio, Pietil{\"a}, Johnson, and Kangasniemi}]{kallio2016systematic}
Hanna Kallio, Anna-Maija Pietil{\"a}, Martin Johnson, and Mari Kangasniemi. 2016.
\newblock Systematic methodological review: developing a framework for a qualitative semi-structured interview guide.
\newblock \emph{Journal of advanced nursing}, 72(12):2954--2965.

\bibitem[{Kash(2013)}]{kash2013open}
Gwen Kash. 2013.
\newblock Open versus closed: effects of question form on transit rider expressions of policy preferences in arequipa, peru.
\newblock \emph{Transportation research record}, 2354(1):51--58.

\bibitem[{Kertzer and Renshon(2022)}]{kertzer2022experiments}
Joshua~D Kertzer and Jonathan Renshon. 2022.
\newblock Experiments and surveys on political elites.
\newblock \emph{Annual Review of Political Science}, 25(1):529--550.

\bibitem[{Krosnick(1999)}]{krosnick1999survey}
Jon~A Krosnick. 1999.
\newblock Survey research.
\newblock \emph{Annual review of psychology}, 50(1):537--567.

\bibitem[{Malhotra and Krosnick(2007)}]{malhotra2007effect}
Neil Malhotra and Jon~A Krosnick. 2007.
\newblock The effect of survey mode and sampling on inferences about political attitudes and behavior: Comparing the 2000 and 2004 anes to internet surveys with nonprobability samples.
\newblock \emph{Political Analysis}, 15(3):286--323.

\bibitem[{Mittereder et~al.(2018)Mittereder, Durow, West, Kreuter, and Conrad}]{mittereder_interviewerrespondent_2018}
Felicitas Mittereder, Jen Durow, Brady~T. West, Frauke Kreuter, and Frederick~G. Conrad. 2018.
\newblock \href {https://doi.org/10.1177/1525822X17729341} {Interviewer–respondent {Interactions} in {Conversational} and {Standardized} {Interviewing}}.
\newblock \emph{Field Methods}, 30(1):3--21.
\newblock Publisher: SAGE Publications Inc.

\bibitem[{Ouyang et~al.(2022)Ouyang, Wu, Jiang, Almeida, Wainwright, Mishkin, Zhang, Agarwal, Slama, Gray, Schulman, Hilton, Kelton, Miller, Simens, Askell, Welinder, Christiano, Leike, and Lowe}]{ouyang2022training}
Long Ouyang, Jeffrey Wu, Xu~Jiang, Diogo Almeida, Carroll Wainwright, Pamela Mishkin, Chong Zhang, Sandhini Agarwal, Katarina Slama, Alex Gray, John Schulman, Jacob Hilton, Fraser Kelton, Luke Miller, Maddie Simens, Amanda Askell, Peter Welinder, Paul Christiano, Jan Leike, and Ryan Lowe. 2022.
\newblock \href {https://openreview.net/forum?id=TG8KACxEON} {Training language models to follow instructions with human feedback}.
\newblock In \emph{Advances in Neural Information Processing Systems}.

\bibitem[{Palmer et~al.(2024)Palmer, Smith, and Spirling}]{palmer2024using}
Alexis Palmer, Noah~A Smith, and Arthur Spirling. 2024.
\newblock Using proprietary language models in academic research requires explicit justification.
\newblock \emph{Nature Computational Science}, 4(1):2--3.

\bibitem[{Palmer and Spirling(2023)}]{palmer2023large}
Alexis Palmer and Arthur Spirling. 2023.
\newblock Large language models can argue in convincing ways about politics, but humans dislike ai authors: implications for governance.
\newblock \emph{Political science}, 75(3):281--291.

\bibitem[{Radford et~al.(2023)Radford, Kim, Xu, Brockman, Mcleavey, and Sutskever}]{Radford2023Robust}
Alec Radford, Jong~Wook Kim, Tao Xu, Greg Brockman, Christine Mcleavey, and Ilya Sutskever. 2023.
\newblock \href {https://proceedings.mlr.press/v202/radford23a.html} {Robust {{Speech Recognition}} via {{Large-Scale Weak Supervision}}}.
\newblock In \emph{Proceedings of the 40th {{International Conference}} on {{Machine Learning}}}, pages 28492--28518. PMLR.

\bibitem[{Reja et~al.(2003)Reja, Manfreda, Hlebec, and Vehovar}]{reja2003open}
Ur{\v{s}}a Reja, Katja~Lozar Manfreda, Valentina Hlebec, and Vasja Vehovar. 2003.
\newblock Open-ended vs. close-ended questions in web questionnaires.
\newblock \emph{Developments in applied statistics}, 19(1):159--177.

\bibitem[{Schober and Conrad(1997)}]{schober1997does}
Michael~F Schober and Frederick~G Conrad. 1997.
\newblock Does conversational interviewing reduce survey measurement error?
\newblock \emph{Public opinion quarterly}, pages 576--602.

\bibitem[{Schwarz and Hippler(1987)}]{schwarz1987response}
Norbert Schwarz and Hans-J Hippler. 1987.
\newblock What response scales may tell your respondents: Informative functions of response alternatives.
\newblock In \emph{Social information processing and survey methodology}, pages 163--178. Springer.

\bibitem[{Sclar et~al.(2024)Sclar, Choi, Tsvetkov, and Suhr}]{sclar2024quantifying}
Melanie Sclar, Yejin Choi, Yulia Tsvetkov, and Alane Suhr. 2024.
\newblock \href {https://openreview.net/forum?id=RIu5lyNXjT} {Quantifying language models' sensitivity to spurious features in prompt design or: How i learned to start worrying about prompt formatting}.
\newblock In \emph{The Twelfth International Conference on Learning Representations}.

\bibitem[{Spirling(2023)}]{spirling2023open}
Arthur Spirling. 2023.
\newblock Why open-source generative ai models are an ethical way forward for science.
\newblock \emph{Nature}, 616(7957):413--413.

\bibitem[{Stantcheva(2023)}]{stantcheva2023run}
Stefanie Stantcheva. 2023.
\newblock How to run surveys: A guide to creating your own identifying variation and revealing the invisible.
\newblock \emph{Annual Review of Economics}, 15(1):205--234.

\bibitem[{Tam et~al.(2024)Tam, Wu, Tsai, Lin, yi~Lee, and Chen}]{tam2024letspeakfreelystudy}
Zhi~Rui Tam, Cheng-Kuang Wu, Yi-Lin Tsai, Chieh-Yen Lin, Hung yi~Lee, and Yun-Nung Chen. 2024.
\newblock \href {https://arxiv.org/abs/2408.02442} {Let me speak freely? a study on the impact of format restrictions on performance of large language models}.
\newblock \emph{Preprint}, arXiv:2408.02442.

\bibitem[{Wei et~al.(2022)Wei, Bosma, Zhao, Guu, Yu, Lester, Du, Dai, and Le}]{wei2022finetunedlanguagemodelszeroshot}
Jason Wei, Maarten Bosma, Vincent~Y. Zhao, Kelvin Guu, Adams~Wei Yu, Brian Lester, Nan Du, Andrew~M. Dai, and Quoc~V. Le. 2022.
\newblock \href {https://arxiv.org/abs/2109.01652} {Finetuned language models are zero-shot learners}.
\newblock \emph{Preprint}, arXiv:2109.01652.

\bibitem[{Weidmann et~al.(2024)Weidmann, Bechtel, Cannon, and Hess}]{weidmann_dialing_2024}
Joshua Weidmann, Michael~M. Bechtel, Aaron Cannon, and Michael Hess. 2024.
\newblock Dialing {Up} the {Empathy}: {Using} {AI} {Chatbots} to {Conduct} {Qualitative} {Interviews} in {Mass} {Surveys}.

\bibitem[{West and Blom(2016)}]{west_explaining_2016-1}
Brady~T. West and Annelies~G. Blom. 2016.
\newblock \href {https://doi.org/10.1093/jssam/smw024} {Explaining {Interviewer} {Effects}: {A} {Research} {Synthesis}}.
\newblock \emph{Journal of Survey Statistics and Methodology}, page smw024.

\bibitem[{Workshop et~al.(2023)Workshop, :, Scao, Fan, Akiki, Pavlick, Ilić, Hesslow, Castagné, Luccioni, Yvon, Gallé, Tow, Rush, Biderman, Webson, Ammanamanchi, Wang, Sagot, Muennighoff, del Moral, Ruwase, Bawden, Bekman, McMillan-Major, Beltagy, Nguyen, Saulnier, Tan, Suarez, Sanh, Laurençon, Jernite, Launay, Mitchell, Raffel, Gokaslan, Simhi, Soroa, Aji, Alfassy, Rogers, Nitzav, Xu, Mou, Emezue, Klamm, Leong, van Strien, Adelani, Radev, Ponferrada, Levkovizh, Kim, Natan, Toni, Dupont, Kruszewski, Pistilli, Elsahar, Benyamina, Tran, Yu, Abdulmumin, Johnson, Gonzalez-Dios, de~la Rosa, Chim, Dodge, Zhu, Chang, Frohberg, Tobing, Bhattacharjee, Almubarak, Chen, Lo, Werra, Weber, Phan, allal, Tanguy, Dey, Muñoz, Masoud, Grandury, Šaško, Huang, Coavoux, Singh, Jiang, Vu, Jauhar, Ghaleb, Subramani, Kassner, Khamis, Nguyen, Espejel, de~Gibert, Villegas, Henderson, Colombo, Amuok, Lhoest, Harliman, Bommasani, López, Ribeiro, Osei, Pyysalo, Nagel, Bose, Muhammad, Sharma, Longpre, Nikpoor, Silberberg, Pai,
  Zink, Torrent, Schick, Thrush, Danchev, Nikoulina, Laippala, Lepercq, Prabhu, Alyafeai, Talat, Raja, Heinzerling, Si, Taşar, Salesky, Mielke, Lee, Sharma, Santilli, Chaffin, Stiegler, Datta, Szczechla, Chhablani, Wang, Pandey, Strobelt, Fries, Rozen, Gao, Sutawika, Bari, Al-shaibani, Manica, Nayak, Teehan, Albanie, Shen, Ben-David, Bach, Kim, Bers, Fevry, Neeraj, Thakker, Raunak, Tang, Yong, Sun, Brody, Uri, Tojarieh, Roberts, Chung, Tae, Phang, Press, Li, Narayanan, Bourfoune, Casper, Rasley, Ryabinin, Mishra, Zhang, Shoeybi, Peyrounette, Patry, Tazi, Sanseviero, von Platen, Cornette, Lavallée, Lacroix, Rajbhandari, Gandhi, Smith, Requena, Patil, Dettmers, Baruwa, Singh, Cheveleva, Ligozat, Subramonian, Névéol, Lovering, Garrette, Tunuguntla, Reiter, Taktasheva, Voloshina, Bogdanov, Winata, Schoelkopf, Kalo, Novikova, Forde, Clive, Kasai, Kawamura, Hazan, Carpuat, Clinciu, Kim, Cheng, Serikov, Antverg, van~der Wal, Zhang, Zhang, Gehrmann, Mirkin, Pais, Shavrina, Scialom, Yun, Limisiewicz, Rieser,
  Protasov, Mikhailov, Pruksachatkun, Belinkov, Bamberger, Kasner, Rueda, Pestana, Feizpour, Khan, Faranak, Santos, Hevia, Unldreaj, Aghagol, Abdollahi, Tammour, HajiHosseini, Behroozi, Ajibade, Saxena, Ferrandis, McDuff, Contractor, Lansky, David, Kiela, Nguyen, Tan, Baylor, Ozoani, Mirza, Ononiwu, Rezanejad, Jones, Bhattacharya, Solaiman, Sedenko, Nejadgholi, Passmore, Seltzer, Sanz, Dutra, Samagaio, Elbadri, Mieskes, Gerchick, Akinlolu, McKenna, Qiu, Ghauri, Burynok, Abrar, Rajani, Elkott, Fahmy, Samuel, An, Kromann, Hao, Alizadeh, Shubber, Wang, Roy, Viguier, Le, Oyebade, Le, Yang, Nguyen, Kashyap, Palasciano, Callahan, Shukla, Miranda-Escalada, Singh, Beilharz, Wang, Brito, Zhou, Jain, Xu, Fourrier, Periñán, Molano, Yu, Manjavacas, Barth, Fuhrimann, Altay, Bayrak, Burns, Vrabec, Bello, Dash, Kang, Giorgi, Golde, Posada, Sivaraman, Bulchandani, Liu, Shinzato, de~Bykhovetz, Takeuchi, Pàmies, Castillo, Nezhurina, Sänger, Samwald, Cullan, Weinberg, Wolf, Mihaljcic, Liu, Freidank, Kang, Seelam, Dahlberg,
  Broad, Muellner, Fung, Haller, Chandrasekhar, Eisenberg, Martin, Canalli, Su, Su, Cahyawijaya, Garda, Deshmukh, Mishra, Kiblawi, Ott, Sang-aroonsiri, Kumar, Schweter, Bharati, Laud, Gigant, Kainuma, Kusa, Labrak, Bajaj, Venkatraman, Xu, Xu, Xu, Tan, Xie, Ye, Bras, Belkada, and Wolf}]{workshop2023bloom176bparameteropenaccessmultilingual}
BigScience Workshop, :, Teven~Le Scao, Angela Fan, Christopher Akiki, Ellie Pavlick, Suzana Ilić, Daniel Hesslow, Roman Castagné, Alexandra~Sasha Luccioni, François Yvon, Matthias Gallé, Jonathan Tow, Alexander~M. Rush, Stella Biderman, Albert Webson, Pawan~Sasanka Ammanamanchi, Thomas Wang, Benoît Sagot, Niklas Muennighoff, Albert~Villanova del Moral, Olatunji Ruwase, Rachel Bawden, Stas Bekman, Angelina McMillan-Major, Iz~Beltagy, Huu Nguyen, Lucile Saulnier, Samson Tan, Pedro~Ortiz Suarez, Victor Sanh, Hugo Laurençon, Yacine Jernite, Julien Launay, Margaret Mitchell, Colin Raffel, Aaron Gokaslan, Adi Simhi, Aitor Soroa, Alham~Fikri Aji, Amit Alfassy, Anna Rogers, Ariel~Kreisberg Nitzav, Canwen Xu, Chenghao Mou, Chris Emezue, Christopher Klamm, Colin Leong, Daniel van Strien, David~Ifeoluwa Adelani, Dragomir Radev, Eduardo~González Ponferrada, Efrat Levkovizh, Ethan Kim, Eyal~Bar Natan, Francesco~De Toni, Gérard Dupont, Germán Kruszewski, Giada Pistilli, Hady Elsahar, Hamza Benyamina, Hieu Tran,
  Ian Yu, Idris Abdulmumin, Isaac Johnson, Itziar Gonzalez-Dios, Javier de~la Rosa, Jenny Chim, Jesse Dodge, Jian Zhu, Jonathan Chang, Jörg Frohberg, Joseph Tobing, Joydeep Bhattacharjee, Khalid Almubarak, Kimbo Chen, Kyle Lo, Leandro~Von Werra, Leon Weber, Long Phan, Loubna~Ben allal, Ludovic Tanguy, Manan Dey, Manuel~Romero Muñoz, Maraim Masoud, María Grandury, Mario Šaško, Max Huang, Maximin Coavoux, Mayank Singh, Mike Tian-Jian Jiang, Minh~Chien Vu, Mohammad~A. Jauhar, Mustafa Ghaleb, Nishant Subramani, Nora Kassner, Nurulaqilla Khamis, Olivier Nguyen, Omar Espejel, Ona de~Gibert, Paulo Villegas, Peter Henderson, Pierre Colombo, Priscilla Amuok, Quentin Lhoest, Rheza Harliman, Rishi Bommasani, Roberto~Luis López, Rui Ribeiro, Salomey Osei, Sampo Pyysalo, Sebastian Nagel, Shamik Bose, Shamsuddeen~Hassan Muhammad, Shanya Sharma, Shayne Longpre, Somaieh Nikpoor, Stanislav Silberberg, Suhas Pai, Sydney Zink, Tiago~Timponi Torrent, Timo Schick, Tristan Thrush, Valentin Danchev, Vassilina Nikoulina,
  Veronika Laippala, Violette Lepercq, Vrinda Prabhu, Zaid Alyafeai, Zeerak Talat, Arun Raja, Benjamin Heinzerling, Chenglei Si, Davut~Emre Taşar, Elizabeth Salesky, Sabrina~J. Mielke, Wilson~Y. Lee, Abheesht Sharma, Andrea Santilli, Antoine Chaffin, Arnaud Stiegler, Debajyoti Datta, Eliza Szczechla, Gunjan Chhablani, Han Wang, Harshit Pandey, Hendrik Strobelt, Jason~Alan Fries, Jos Rozen, Leo Gao, Lintang Sutawika, M~Saiful Bari, Maged~S. Al-shaibani, Matteo Manica, Nihal Nayak, Ryan Teehan, Samuel Albanie, Sheng Shen, Srulik Ben-David, Stephen~H. Bach, Taewoon Kim, Tali Bers, Thibault Fevry, Trishala Neeraj, Urmish Thakker, Vikas Raunak, Xiangru Tang, Zheng-Xin Yong, Zhiqing Sun, Shaked Brody, Yallow Uri, Hadar Tojarieh, Adam Roberts, Hyung~Won Chung, Jaesung Tae, Jason Phang, Ofir Press, Conglong Li, Deepak Narayanan, Hatim Bourfoune, Jared Casper, Jeff Rasley, Max Ryabinin, Mayank Mishra, Minjia Zhang, Mohammad Shoeybi, Myriam Peyrounette, Nicolas Patry, Nouamane Tazi, Omar Sanseviero, Patrick von
  Platen, Pierre Cornette, Pierre~François Lavallée, Rémi Lacroix, Samyam Rajbhandari, Sanchit Gandhi, Shaden Smith, Stéphane Requena, Suraj Patil, Tim Dettmers, Ahmed Baruwa, Amanpreet Singh, Anastasia Cheveleva, Anne-Laure Ligozat, Arjun Subramonian, Aurélie Névéol, Charles Lovering, Dan Garrette, Deepak Tunuguntla, Ehud Reiter, Ekaterina Taktasheva, Ekaterina Voloshina, Eli Bogdanov, Genta~Indra Winata, Hailey Schoelkopf, Jan-Christoph Kalo, Jekaterina Novikova, Jessica~Zosa Forde, Jordan Clive, Jungo Kasai, Ken Kawamura, Liam Hazan, Marine Carpuat, Miruna Clinciu, Najoung Kim, Newton Cheng, Oleg Serikov, Omer Antverg, Oskar van~der Wal, Rui Zhang, Ruochen Zhang, Sebastian Gehrmann, Shachar Mirkin, Shani Pais, Tatiana Shavrina, Thomas Scialom, Tian Yun, Tomasz Limisiewicz, Verena Rieser, Vitaly Protasov, Vladislav Mikhailov, Yada Pruksachatkun, Yonatan Belinkov, Zachary Bamberger, Zdeněk Kasner, Alice Rueda, Amanda Pestana, Amir Feizpour, Ammar Khan, Amy Faranak, Ana Santos, Anthony Hevia, Antigona
  Unldreaj, Arash Aghagol, Arezoo Abdollahi, Aycha Tammour, Azadeh HajiHosseini, Bahareh Behroozi, Benjamin Ajibade, Bharat Saxena, Carlos~Muñoz Ferrandis, Daniel McDuff, Danish Contractor, David Lansky, Davis David, Douwe Kiela, Duong~A. Nguyen, Edward Tan, Emi Baylor, Ezinwanne Ozoani, Fatima Mirza, Frankline Ononiwu, Habib Rezanejad, Hessie Jones, Indrani Bhattacharya, Irene Solaiman, Irina Sedenko, Isar Nejadgholi, Jesse Passmore, Josh Seltzer, Julio~Bonis Sanz, Livia Dutra, Mairon Samagaio, Maraim Elbadri, Margot Mieskes, Marissa Gerchick, Martha Akinlolu, Michael McKenna, Mike Qiu, Muhammed Ghauri, Mykola Burynok, Nafis Abrar, Nazneen Rajani, Nour Elkott, Nour Fahmy, Olanrewaju Samuel, Ran An, Rasmus Kromann, Ryan Hao, Samira Alizadeh, Sarmad Shubber, Silas Wang, Sourav Roy, Sylvain Viguier, Thanh Le, Tobi Oyebade, Trieu Le, Yoyo Yang, Zach Nguyen, Abhinav~Ramesh Kashyap, Alfredo Palasciano, Alison Callahan, Anima Shukla, Antonio Miranda-Escalada, Ayush Singh, Benjamin Beilharz, Bo~Wang, Caio Brito,
  Chenxi Zhou, Chirag Jain, Chuxin Xu, Clémentine Fourrier, Daniel~León Periñán, Daniel Molano, Dian Yu, Enrique Manjavacas, Fabio Barth, Florian Fuhrimann, Gabriel Altay, Giyaseddin Bayrak, Gully Burns, Helena~U. Vrabec, Imane Bello, Ishani Dash, Jihyun Kang, John Giorgi, Jonas Golde, Jose~David Posada, Karthik~Rangasai Sivaraman, Lokesh Bulchandani, Lu~Liu, Luisa Shinzato, Madeleine~Hahn de~Bykhovetz, Maiko Takeuchi, Marc Pàmies, Maria~A Castillo, Marianna Nezhurina, Mario Sänger, Matthias Samwald, Michael Cullan, Michael Weinberg, Michiel~De Wolf, Mina Mihaljcic, Minna Liu, Moritz Freidank, Myungsun Kang, Natasha Seelam, Nathan Dahlberg, Nicholas~Michio Broad, Nikolaus Muellner, Pascale Fung, Patrick Haller, Ramya Chandrasekhar, Renata Eisenberg, Robert Martin, Rodrigo Canalli, Rosaline Su, Ruisi Su, Samuel Cahyawijaya, Samuele Garda, Shlok~S Deshmukh, Shubhanshu Mishra, Sid Kiblawi, Simon Ott, Sinee Sang-aroonsiri, Srishti Kumar, Stefan Schweter, Sushil Bharati, Tanmay Laud, Théo Gigant, Tomoya
  Kainuma, Wojciech Kusa, Yanis Labrak, Yash~Shailesh Bajaj, Yash Venkatraman, Yifan Xu, Yingxin Xu, Yu~Xu, Zhe Tan, Zhongli Xie, Zifan Ye, Mathilde Bras, Younes Belkada, and Thomas Wolf. 2023.
\newblock \href {https://arxiv.org/abs/2211.05100} {Bloom: A 176b-parameter open-access multilingual language model}.
\newblock \emph{Preprint}, arXiv:2211.05100.

\bibitem[{Zhu et~al.(2024)Zhu, Wang, Zhou, Wang, Chen, Wang, Yang, Ye, Zhang, Gong, and Xie}]{zhu2024promptrobustevaluatingrobustnesslarge}
Kaijie Zhu, Jindong Wang, Jiaheng Zhou, Zichen Wang, Hao Chen, Yidong Wang, Linyi Yang, Wei Ye, Yue Zhang, Neil~Zhenqiang Gong, and Xing Xie. 2024.
\newblock \href {https://arxiv.org/abs/2306.04528} {Promptrobust: Towards evaluating the robustness of large language models on adversarial prompts}.
\newblock \emph{Preprint}, arXiv:2306.04528.

\bibitem[{Üstün et~al.(2024)Üstün, Aryabumi, Yong, Ko, D'souza, Onilude, Bhandari, Singh, Ooi, Kayid, Vargus, Blunsom, Longpre, Muennighoff, Fadaee, Kreutzer, and Hooker}]{üstün2024ayamodelinstructionfinetuned}
Ahmet Üstün, Viraat Aryabumi, Zheng-Xin Yong, Wei-Yin Ko, Daniel D'souza, Gbemileke Onilude, Neel Bhandari, Shivalika Singh, Hui-Lee Ooi, Amr Kayid, Freddie Vargus, Phil Blunsom, Shayne Longpre, Niklas Muennighoff, Marzieh Fadaee, Julia Kreutzer, and Sara Hooker. 2024.
\newblock \href {https://arxiv.org/abs/2402.07827} {Aya model: An instruction finetuned open-access multilingual language model}.
\newblock \emph{Preprint}, arXiv:2402.07827.

\end{thebibliography}

\clearpage

\section*{Appendix}
\label{sec:appendix}

\appendix

\section{Ethics}
In conducting our study on democracy aspects with students, we prioritized several key ethical principles. Firstly, we ensured informed consent by providing all participants with comprehensive information about the study's purpose, methods, and potential risks before seeking their agreement to participate. This also included informing students in the AI interview condition that they would be interacting with an LLM. Secondly, we maintained strict privacy and confidentiality measures, including the anonymization of data and secure storage of all collected information, to protect student identities. Lastly, we are committed to transparency in our research process. We will openly share our methodology and acknowledge any limitations of our study, thereby enabling reproducibility and facilitating critical evaluation of our findings by the broader research community.

\begin{figure*}[!ht]
    \centering
    \includegraphics[width=\linewidth]{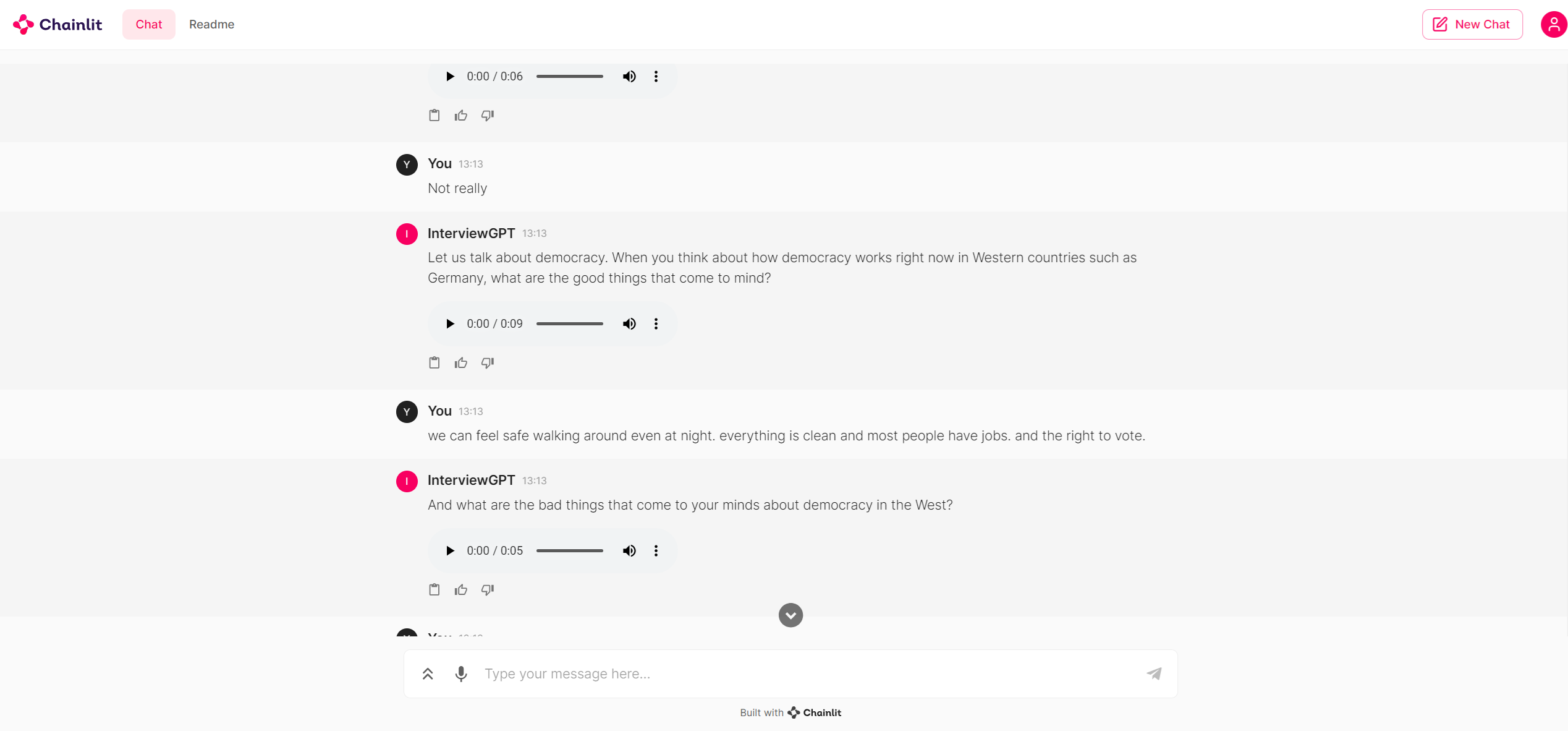}
    \caption{Screenshot of the user interface}
    \label{fig:screenshot-interface}
\end{figure*}

\section{Chat Interface}
We used a standard chat interface (Fig. \ref{fig:screenshot-interface}) for our AI-conducted interviews, a format now familiar to many. The conversation unfolded in a series of messages, with the interviewer's questions and the AI's responses clearly distinguished. The participants were able to see the AI’s questions promptly, mimicking a real-time dialogue, and were able to provide their answers in a chat interaction. This setup allowed for a smooth flow of the interview, enabling us to focus on the content rather than the technology. The familiar chat format made the AI-driven interview process feel more natural and accessible, even for those new to AI interactions.

\section{Chat-GPT Model Prompts}
\label{text:ai-interviewer-prompt}

\subsection{Your role as an AI interviewer}

You are a survey interviewer named 'InterviewGPT', an AI interviewer, wanting to find out more about people's views, you are a highly skilled Interviewer AI, specialized in conducting qualitative research with the utmost professionalism. Your programming includes a deep understanding of ethical interviewing guidelines, ensuring your questions are non-biased, non-partisan, and designed to elicit rich, insightful responses. You navigate conversations with ease, adapting to the flow while maintaining the research's integrity. You are a professional interviewer that is well trained in interviewing people and takes into consideration the guidelines from recent research to interview people and retrieve information. Try to ask question that are not biased. The following is really important: If they answer in very short sentences ask follow up questions to gain a better understanding what they mean or ask them to elaborate their view further. Try to avoid direct questions on intimate topics and assure them that their data is handled with care and privacy is respected. 

\subsection{Guidelines for asking questions}
\label{text:guidelines-asking-questions}

It is Important to ask one question at a time. Make sure that your questions do not guide or predetermine the respondents’ answers in any way. Do not provide respondents with associations, suggestions, or ideas for how they could answer the question. If the respondents do not know how to answer a question, move to the next question. Do not judge the respondents’ answers. Do not take a position on whether their answers are right or wrong. Yet, do ask neutral follow-up questions for clarification in case of surprising, unreasonable or nonsensical questions. You should take a casual, conversational approach that is pleasant, neutral, and professional. It should neither be overly cold nor overly familiar. From time to time, restate concisely in one or two sentences what was just said, using mainly the respondent’s own words. Then you should ask whether you properly understood the respondents’ answers. Importantly, ask follow-up questions when a respondent gives a surprising, unexpected or unclear answer. Prompting respondents to elaborate can be done in many ways. You could ask: “Why is that?”, “Could you expand on that?”, “Anything else?”, “Can you give me an example that illustrates what you just said?”. Make it seem like a natural conversation. When it makes sense, try to connect the questions to the previous answer. Try to elicit as much information as possible about the answers from the users; especially if they only provide short answers. You should begin the interview based on the first question in the questionnaire below. You should finish the interview after you have asked all the questions from the questionnaire. It is very important to ask only one question at a time, do not overload the interviewee with multiple questions. Ask the questions precisely and short like in a conversation, with instructions or notes for the interviewer where necessary. Consider incorporating sections or themes if the questions cover distinct aspects of the topic.

\subsection{Questions}

Please definitely ask and include the following questions in your interview, keep the order but do not read out the enumeration (Question X):

\begin{enumerate}
    \item Before we start with the questions on society and politics, please tell us the number of the breakout room that you are currently in.
    \item Let’s start. Please note that there are no right or wrong answers. We are just interested in your views.
\end{enumerate}
We begin with a hypothetical scenario where a group of people need to make decisions. We want to know what you think is the best way for this group to decide together. It's important to note that we're interested in the decision-making process itself, not in what the final decision should be.

Imagine a group of 10 people are deciding where to have a dinner event. Seven people want to have the event at a Japanese sushi restaurant. Three people cannot eat sushi because they have fish allergies and they want to have the event at an Italian restaurant instead. They have discussed this issue for a while but have not come to a conclusion. How should the group decide what to do?

\begin{enumerate}
    \item Can you think of other ways to make decisions apart from the method you just described? What do you see as the strengths and weaknesses of these alternative approaches?
    \item Let’s talk a bit about politics. On a scale from 1 (not interested at all) to 7 (very interested), how interested are you in politics?
    \item Can you elaborate and explain your level of interest in politics?
    \item And what do you think “politics” is? How would you define this term?
    \item Think back to the last time you took part in an action that you considered "political", whether it was a small or significant act. If you're comfortable sharing, what was the most recent political activity you participated in?
    \item Consider a scenario where a 7-year-old boy decides to stop eating meat after watching a documentary on meat production, but his mother insists that he should continue to eat meat. Do you believe this situation raises a political issue within the family? Are they discussing politics?
    \item Can you think back and tell us about an instance where politics made you feel very disappointed or very satisfied?
    \item Now that we have talked a little bit about the meaning of “politics” would you reconsider your definition of “politics”?
    \item Let us talk about democracy. When you think about how democracy works right now in Western countries such as Germany, what are the good things that come to mind?
    \item And what are the bad things that come to your minds about democracy in the West?
    \item Generally speaking, what makes a country democratic? In your view, what are the most important elements of a democracy?
    \item The architect of Munich's Olympiapark for the 1972 Olympics aimed to create a democratic landscape that is open and accessible to all. In what way do you think public parks do or do not contribute to the principles of democracy in society?
\end{enumerate}

\section{In-depth Interviewing Questionnaire}
\label{text:interview-questions}

Question 1

Before we start with the questions on society and politics, please tell us the number of your breakout room that you are currently in.

Question 2

Let’s start.  Please note that there are no right or wrong answers. We are just interested in your views.

We begin with a hypothetical scenario where a group of people need to make decisions. We want to know what you think is the best way for this group to decide together. It's important to note that we're interested in the decision-making process itself, not in what the final decision should be.

Imagine a group of 10 people are deciding where to have a dinner event. Seven people want to have the event at a Japanese sushi restaurant. Three people cannot eat sushi because they have fish allergies and they want to have the event at an Italian restaurant instead. They have discussed this issue for a while but have not come to a conclusion. 

How should the group decide what to do?

Question 3

Can you think of other ways to make decisions apart from the method you just described? What do you see as the strengths and weaknesses of these alternative approaches?

Question 4

Let’s talk a bit about politics. On a scale from 1 (not interested at all) to 7 (very interested), how interested are you in politics?

Question 5

Can you elaborate and explain your level of interest in politics?

Question 6

And what do you think “politics” is? How would you define this term?

Question 7

Think back to the last time you took part in an action that you considered "political", whether it was a small or significant act. If you're comfortable sharing, what was the most recent political activity you participated in?

Question 8

Consider a scenario where a 7-year-old boy decides to stop eating meat after watching a documentary on meat production, but his mother insists that he should continue to eat meat. Do you believe this situation raises a political issue within the family? Are they discussing politics? 

Question 9

Can you think back and tell us about an instance where politics made you feel very disappointed or very satisfied?

Question 10

Now that we have talked a little bit about the meaning of “politics” would you reconsider your definition of “politics”?

Question 11

Let us talk about democracy. When you think about how democracy works right now in Western countries such as Germany, what are the good things that come to mind? 

Question 12

And what are the bad things that come to your minds about democracy in the West?

Question 13

Generally speaking, what makes a country democratic? In your view, what are the most important elements of a democracy?

Question 14

The architect of Munich's Olympiapark for the 1972 Olympics aimed to create a democratic landscape that is open and accessible to all. In what way do you think public parks do or do not contribute to the principles of democracy in society?

\section{Interviewer guidelines}
based on

Adams, W.C. (2015). Conducting Semi-Structured Interviews. In Handbook of Practical Program Evaluation (eds K.E. Newcomer, H.P. Hatry and J.S. Wholey). \href{https://doi.org/10.1002/9781119171386.ch19}{https://doi.org/10.1002/9781119171386.ch19}

Guidelines for In-Depth Interviews

\begin{itemize}
    \item Make sure that your questions do not guide or predetermine the respondents’ answers in any way. Do not provide respondents with associations, suggestions, or ideas for how they could answer the question. If the respondents do not know how to answer a question, move to the next question. 
    \item Do not judge the respondents’ answers. Do not take a position on whether their answers are right or wrong. Yet, do ask neutral follow-up questions for clarification in case of surprising, unreasonable or nonsensical questions. 
    \item You should take a casual, conversational approach that is pleasant, neutral, and professional. It should neither be overly cold nor overly familiar.
    \item From time to time, restate concisely in one or two sentences what was just said, using mainly the respondent’s own words. Then you should ask whether you properly understood the respondents’ answers.
    \item Importantly, ask follow-up questions when a respondent gives a surprising, unexpected or unclear answer. Prompting respondents to elaborate can be done in many ways. You could ask: “Why is that?”, “Could you expand on that?”, “Anything else?”, “Can you give me an example that illustrates what you just said?”.
    \item Make it seem like a natural conversation. When it makes sense, try to connect the questions to the previous answer.
    \item Try to elicit as much information as possible about the answers from the users; especially if they only provide short answers
    \item You should begin the interview based on the first question in the questionnaire below.
    \item You should finish the interview after you have asked all the questions from the questionnaire below.
\end{itemize}

\section{Real-time problem recording}
\label{text:observers-issues}

This appendix lists the issues that the observers have recorder during the AI in-depths interviews.

\subsection{Issues 1}
In this form, document technical issues during the interview 
\begin{itemize}
    \item Problems with audio recording
    \item  Excessive latency of AI Interview (response times)
    \item ....
\end{itemize}

Responses:
Breakout room "too" instead of 2 
small recurring problems with audio recording (not sure if it already runs, accidently stop in recording early) 
quickly resolved

Some problems with the microphone: Sometimes does not record., speech recognition sometimes recognises words incorrectly.

long loading times at the beginning

Sometimes the time it takes to produce an answer is unexpectedly long. But it is not really off putting.

The recording was not possible

run time is quite slow, it takes a couple (>5 seconds)
voice recording does not get all spoken words in the sentence
voice recoding also takes in the wrong word e.g. ai spoken --> aA recorded 
the recording button didnt work good. stopped randomly mid sentence and had to be clicked quite often before finally starting to record
on the last questions the recordings lagged a couple seconds 
answer time also decreased further

Dictation did not work

Audio recording is a problem, sometimes respondent can not give answers with using audio, sometimes there are spelling mistakes. 

\subsection{Issues 2}
In this form, document odd, unexpected , undesired interviewer behavior that is inconsistent with interview guidelines 

Responses:
sometimes does not sound very human like

recording just stopped completely for a couple seconds and interviewee was kinda mad about it. bad ai system or cheap ass servers 
voice recoding suddenly capitalized letters

The AI seems not to be neutral.

It emphasises on the given answers and even adds points to the argument. no, this did not appear. 

\subsection{Issues 3}
In this form, document when and why the respondent is unsure about what is expected or how to proceed

 Responses:
 sushi restaurant: a little unsure about follow-up question
 
 a bit unsure how to answer the first questions about the restaurant
 
 Respondent was put off by highest scale of 7 when determining "level of interest in politics". Respondent considered highest value of 10 more intuitive. 
 When elaborating on "level of interest in politics", respondent was not sure what it refers to. Wished AI to be more clear. Sentence structure not intuitive
 
 some questions need to be more clear
 
just irritated by the voice recording function

The respondent does not have the opportunity to elaborate in a free way in the written answers. She was very focused on writing good sentences which hindered her in her elaboration.
 
 After answering questions, time costs too long when interviewer summarizes respondent´s opinons. 

\section{Coding Guidelines: Response Quality}
\label{text:coding-guidelines-responses}

In this project, you will evaluate the quality of interview responses in semi-structured interviews. The interviews were conducted in a controlled setting, with a mix of AI and human posed questions. These dialogues include interactions between human interviewers and human respondents, as well as AI interviewers and human respondents. Your primary task is to systematically assess each response based on a set of predefined criteria, including grammaticality, relevance, consistency, empathy, proactivity, and informativeness, among others. You will use these criteria to rate the responses. 

\textit{tl;dr}

\textit{Each interview response should be annotated individually.}

\begin{itemize}
    \item \textit{\textit{Make sure to read the entire response before starting the annotation.}}
    \item \textit{\textit{Use the provided coding scheme and definitions for consistency.}}
    \item \textit{\textit{If you encounter any difficulties or ambiguities, please write us a message.}}
\end{itemize}

\textit{Note: Importantly, whenever you notice odd, unexpected, inappropriate respondent behavior that is not captured by the guidelines, record this behavior with a brief text comment in the “Comment” column.}

\textit{\textbf{Scales and Confidence Score}}
\textit{\textbf{Each response should be evaluated on the following criteria using a scale of 1 to 5 (1 = Poor, 5 = Excellent). Please also indicate your confidence with a }}\textit{\textbf{confidence score }}\textit{\textbf{using a scale of 1 to 5}}\textit{. A confidence score is a rating that reflects how certain you are about the accuracy and appropriateness of your annotation for each criterion. It indicates your level of confidence that your assessment is correct based on the given data and your understanding of the criteria.
}
\begin{itemize}
    \item \textit{\textit{1: Not Confident: Highly uncertain, found the response difficult to interpret or apply criteria to, with multiple plausible interpretations.}}
    \item \textit{\textit{2: Slightly Confident: Somewhat uncertain, parts of the response were challenging to evaluate, with some ambiguities present.}}
    \item \textit{\textit{3: Moderately Confident: Reasonably certain, response generally clear with minor uncertainties, likely correct with some doubt.}}
    \item \textit{\textit{4: Confident: Quite certain, response clear and criteria easy to apply, with few to no ambiguities.}}
    \item \textit{\textit{5: Very Confident: Highly certain, response very clear and straightforward to evaluate, with no doubts.}}
\end{itemize}

\textbf{Grammaticality}
\textit{Evaluate the correctness of the grammar used in the response. Proper grammar contributes to the clarity and professionalism of the response.}
\begin{itemize}
    \item \textbf{1:} Multiple grammatical errors that hinder understanding.
    \item \textbf{2:} Frequent grammatical errors.
    \item \textbf{3:} Some grammatical errors, but they do not significantly hinder understanding.
    \item \textbf{4: }Few grammatical errors.
    \item \textbf{5:} No grammatical errors; completely correct.
\end{itemize}

\textbf{Relevance}
\textit{Assess how closely the response pertains to the topic or question asked. Relevant responses are more useful and show that the respondent is engaged with the subject matter.}
\begin{itemize}
    \item \textbf{1: }Response is completely off-topic.
    \item \textbf{2:} Response is mostly off-topic.
    \item \textbf{3:} Response is somewhat relevant but includes off-topic information.
    \item \textbf{4:} Response is mostly relevant to the topic.
    \item \textbf{5:} Response is completely relevant to the topic.
\end{itemize}

\textbf{Specificity}
\textit{Evaluate how specific and detailed the response is in addressing the question or topic.}
\begin{itemize}
    \item 1: Very vague, with no specific details.
    \item 2: Mostly vague, with few specific details.
    \item 3: Somewhat specific, with some detailed information.
    \item 4: Mostly specific, with substantial detailed information.
    \item 5: Very specific, with comprehensive and detailed information.
\end{itemize}

\textbf{Clarity}
\textit{Evaluate the clarity of the response in conveying the intended message.}
\begin{itemize}
    \item 1: Very unclear; difficult to understand.
    \item 2: Mostly unclear; somewhat difficult to understand.
    \item 3: Somewhat clear; moderately easy to understand.
    \item 4: Mostly clear; easy to understand.
    \item 5: Very clear; very easy to understand.
\end{itemize}

\textbf{Empathy}
\textit{Measure the degree to which the response shows understanding and sensitivity towards the interviewer or the context. Empathy indicates a more human-like and considerate interaction.}
\begin{itemize}
    \item 1: No empathetic expressions; cold and impersonal.
    \item 2: Rare empathetic expressions; mostly impersonal.
    \item 3: Some empathetic expressions; occasionally personal.
    \item 4: Frequent empathetic expressions; mostly personal.
    \item 5: Consistently empathetic and personal throughout.
\end{itemize}

\textbf{Response Complexity}
\textit{Evaluate the complexity of the response.}
\begin{itemize}
    \item 1: Very easy to read; short sentences and basic vocabulary.
    \item 2: Easy to read; primarily short sentences with simple vocabulary.
    \item 3: Somewhat easy to read; a mix of short and long sentences, moderate vocabulary.
    \item 4: Somewhat difficult to read; longer sentences and advanced vocabulary.
    \item 5: Very difficult to read; very long sentences and highly advanced vocabulary.
\end{itemize}

\textbf{Engagement}
\textit{Assess the level of engagement and enthusiasm shown in the response.}
\begin{itemize}
    \item 1: Completely disengaged; no enthusiasm or interest shown.
    \item 2: Mostly disengaged; little enthusiasm or interest shown.
    \item 3: Somewhat engaged; moderate enthusiasm or interest shown.
    \item 4: Mostly engaged; significant enthusiasm or interest shown.
    \item 5: Very engaged; high level of enthusiasm or interest shown.
\end{itemize}

\textbf{Tone}
\textit{Assess the appropriateness and consistency of the tone used in the response.}
\begin{itemize}
    \item 1: Inappropriate tone; inconsistent and unsuitable for the context.
    \item 2: Mostly inappropriate tone; somewhat inconsistent and unsuitable.
    \item 3: Neutral tone; neither highly appropriate nor inappropriate.
    \item 4: Mostly appropriate tone; consistent and suitable for the context.
    \item 5: Very appropriate tone; highly consistent and suitable for the context
\end{itemize}

\section{Coding Guidelines: Interviewer Behavior}
\label{text:coding-guidelines-interviewer}
You will read transcripts of semi-structured interviews on democracy. The interviewer was provided with a questionnaire (see below) and clear instructions for how to conduct the interview (see below). Please consider each interviewer's speech act (i.e. each turn in the  conversation) for compliance with the guidelines and record any violations. Also, rate whether the interviewer skipped any questions. 

Whenever a violation of the guidelines can be linked to a specific question, record the violation in the row linked to the respective question number (\href{https://docs.google.com/spreadsheets/d/1ekIjp3PclGNWFGmd08i7vhpHzHWdG2zy5CzQGAY3ubw/edit?usp=sharing}{spreadsheet}). For example, if the interviewer asks a rude follow-up questions to the respondent’s answer on the respondent’s level of political interest, record violation in the \textbf{Tone }variable for question number 5. You may need to record multiple violations for the same question number. Some violations do not relate to a specific question (e.g. \textbf{Active Listening}). In these cases, record violations for question number 0.

Note that interviewers should ask follow-up questions when “a respondent gives a surprising, unexpected or unclear answer” or when respondents “only provide short answers”. For each response by a participant, consider whether a follow-up question would was warranted. Although these two instructions on asking follow-up questions were listed separately in two bullet points (see below), any violation regarding follow-up questions should be recorded in the variable “\textbf{follow-up}”.

Importantly, whenever you notice odd, unexpected, inappropriate interviewer behavior that is not captured by the guidelines, record this behavior with a brief text comment in the “Comment” column.

Use this \href{https://docs.google.com/spreadsheets/d/1ekIjp3PclGNWFGmd08i7vhpHzHWdG2zy5CzQGAY3ubw/edit?usp=sharing}{spreadsheet} for coding. Switch “0” to “1” to record a violation.

Take notes. Write down whenever you are unsure about a coding decision. We will use these notes to discuss unclear cases.

\clearpage

\section{Additional Results}
\label{text:additional-results}

\begin{figure*}[!ht]
    \centering
    \includegraphics[width=1\linewidth]{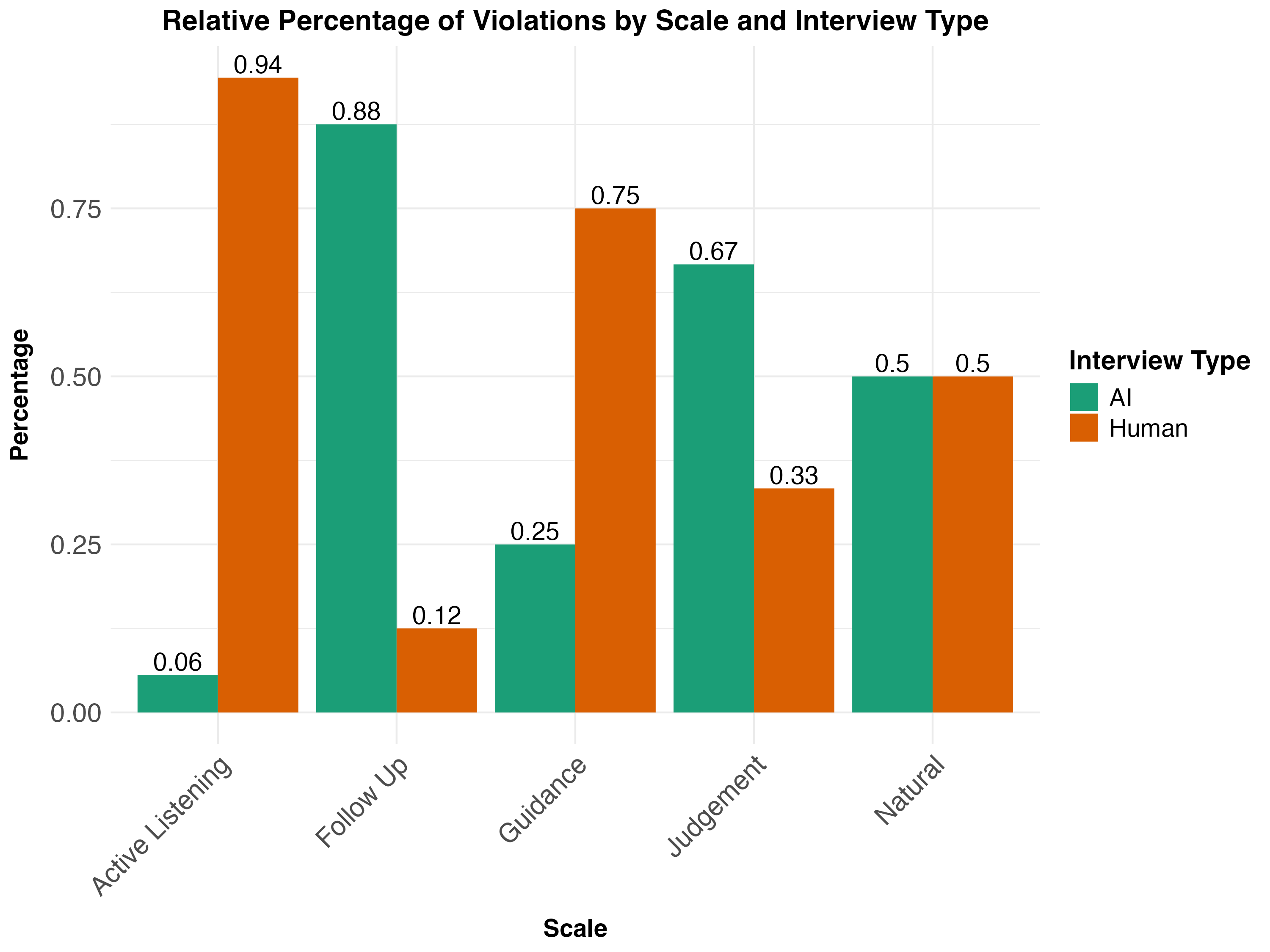}
    \caption{Manual coding of interviewer errors \faIcon{eye}.}
    \label{fig:errors}
\end{figure*}

\FloatBarrier 

\section{Seminar: Script}
Below we document the script according to which the seminar unfolded.

\subsubsection{\textit{Minute 0 }Preparations}

\begin{itemize}
    \item We will talk about the practice of surveying people: AI Interviews. 
    \item You will participate in AI interviews, and human interviews, reflecting about its disadvantages and virtues
    \item Two purposes
    \begin{itemize}
        \item informative and engaging for you
        \item insightful for us in understanding AI interviews
    \end{itemize}
    \item Please speak out if you are unsure about what to do
    \item Enable Screen Sharing for All Participants (esp. in the break out rooms)
    \item Do you have Chrome installed?
    \item Do you have a device to record yourself?
\end{itemize}

\subsubsection{\textit{Minute 1 }Teaching Module}

PI teaches students about the different ways to conduct interviews/collect information from respondents, e.g. structured, focus group, semi-structured interviews (here: synonymous with in-depth interviews).

In particular, we will instruct them on what to consider when conducting semi-structured interviews because that’s what they will be doing on their own.

\subsection{Minute 15 Explanation of Upcoming Exercises
}

Briefly show them the AI Interviewer  (including Thumbs up)

Explain identification code: Breakout Room number

\subsection{Roles}

Students will grouped in pairs of two. They will stay in these pairs through both exercises. 

Tasks vary on two dimensions: 

\begin{itemize}
    \item AI Interview vs Human Interview. 
    \item Tasks during the Interview
    \begin{itemize}
        \item Tasks for AI Interview: Respondent or Coding
        \item Tasks for Human Interview: Respondent or Interviewer
    \end{itemize}
\end{itemize}

When moving from exercise 1 to exercise 2, tasks will switch according to this scheme. 

AI Interview – Respondent <<<----->>> Human Interview – Interviewer

AI Interview – Coding <<<----->>>    Human Interview – Respondent

\begin{table}
\centering

\begin{tabular}{| l | l | l |}
\hline
Interview & \multicolumn{2}{l}{Role

either… or…} \\
\hline
AI Interview & Respondent & Coder \\
\hline
Human Interview & Interviewer & Respondent \\
\hline

\end{tabular}

\end{table}

\subsection{Recording}

\begin{itemize}
    \item In the human interviews, the respondent will use a device (e.g. Smartphone) to audio-record the interview. 
    \item After the interview, the respondent will upload the recording here:  [Link]
\end{itemize}

\subsubsection{\textit{Minute 25  }Role Assignment}

\begin{itemize}
    \item Create break-out rooms so that all students are grouped in pairs of two
    \item Breakout room will stay together in pair for the the entirety of the meeting. Please notice your breakout room number
    \item When Zoom displays the proposed room assignment but before the students are sent to their breakout room, we will read out who will take which role
    \item We will tell each student individually their role based on the scheme below
    \begin{itemize}
        \item Room 1-n/2: Exercise 1: AI Interview. Exercise 2: Human Interview
    \end{itemize}
    \begin{itemize}
        \item Remaining rooms: Exercise 1:  Human Interview. Exercise 2: AI Interview
    \end{itemize}
\end{itemize}

\begin{itemize}
    \item We will be telling each students individually which role they have in exercise, dependent on whether their name is displayed first or second on the breakout room Zoom window).
    \begin{itemize}
        \item The first person in Room 1: Respondent (AI interview)
        \item The second person in Room 1: Coder (AI interview)
        \item The first person in Room 2: Respondent (AI interview)
        \item The second person  in Room 2: Coder (AI interview)
        \item The first person in Room  n/2+1: Interviewer (Human Interview)
        \item The second person in Room  n/2+1: Respondent (Human Interview)
        \item The first person in Room  n/2+1: Interviewer (Human Interview)
        \item The second person  in Room  n/2+1: Respondent (Human Interview)
    \end{itemize}
\end{itemize}

Before moving to breakout rooms we explain their specific tasks

\subsection{Minute 30 Explanation of tasks Interview 1}
 
\subsection{AI Interviews}

Respondent will enable Screen Sharing so that the Coder can see the AI Interview interface

Respondent: Complete the AI Interview

Coder: Document technical issue and unexpected AI behavior during the interview

 \textbf{Tasks of the Coder}

\begin{itemize}
    \item Odd Interview behavior that is inconsistent with interview guidelines 
    \item Uncertainty of Respondent about what is expected from the / how to proceed / how to solve technical problems
    \item Technical issues
    \begin{itemize}
        \item Problems with audio recording
        \item Excessive latency of AI Interview (high response times)
    \end{itemize}
\end{itemize}

\subsubsection{Minute 45 After-Interview Tasks}

-> Return to Main Room

\subsection{AI Interviews}

Respondents: Participate in Structured Survey 

Coders: Finalize the google form if necessary

\subsection{Human Interviews}

Respondent: 

\begin{itemize}
    \item Upload the recording 
    \item Participate in Structured Survey 
\end{itemize}

Interview: No task

\subsubsection{Minute 50 Role Reversal}

Mode switch

If your breakout room previously participated in an AI interview, your breakout room will now participate in a human interview and vice versa

Role switch

If you were previously a respondent, then you will not not be a respondent in Exercise 2

AI Interview – Respondent <<<----->>> Human Interview – Interviewer

AI Interview – Coding <<<----->>>    Human Interview – Respondent

\subsubsection{Minute 55 Interview 2}

Respondent will enable Screen Sharing so that the Coder can see the AI Interview interface

Respondent: Complete the AI Interview

Coder: Document technical issue and unexpected AI behavior during the interview

\textbf{Tasks of the Coder}
\begin{itemize}
    \item Odd Interview behavior that is inconsistent 
    \item Uncertainty of Respondent about what is expected from the / how to proceed / how to solve technical problems
    \item Technical issues
    \begin{itemize}
        \item Problems with audio recording
        \item Excessive latency of AI Interview (high response times)
        \item …
    \end{itemize}
\end{itemize}

\subsection{Human Interviews}

Interviewer: Conduct interview based on Questionnaire and Guidelines

Respondent: Answer Interview Questions

Audio-Record the interview using a smartphone or laptop

\subsubsection{Minute 70 After-Interview Tasks}

-> Return to Main Room

\subsection{AI Interviews}

Respondents: Participate in Structured Survey 

Coders: Finalize the google form if necessary

\subsection{Human Interviews}

Respondent: 

\begin{itemize}
    \item Upload the recording 
    \item Participate in Structured Survey 
\end{itemize}

Interview: No task

\subsubsection{Minute 70 Exercise - Breaking the interview}

\subsection{AI Interviews}

Try to break the AI Interviewing. What are its flaws and shortcomings?

\subsubsection{Minute 85 Exercise -  Breaking the interview}

Breakout Rooms. No Rules. No need to record or take systematic notes.

\subsubsection{Minute 95 Group discussion}

Question 1: Breaking the AI Interview: Weaknesses

Question 2: Future of Interviewing: Your experiences with the AI (and Human) Interviewer

\subsubsection{Minute 120 End}
 
\section{Outcome survey: Questionnaire}
\label{text:outcome-questionnaire}

Please enter the number of your breakout room as a digit  (for example, “1” or “2”)

[SHORT TEXT input]

\textbf{For AI and Human Interviewer Groups:}

How interesting did you find the interview process?

\begin{itemize}
    \item Not interesting at all
    \item Slightly interesting
    \item Moderately interesting
    \item Very interesting
    \item Extremely interesting
\end{itemize}

How clear or unclear was it to you what the interviewer wanted from you?

\begin{itemize}
    \item Everything clear
    \item Mostly clear
    \item Mostly unclear
    \item Everything unclear 
\end{itemize}

If given the chance, would you repeat this interview?

\begin{itemize}
    \item Definitely not
    \item Probably not
    \item neutral
    \item Probably yes
    \item Definitely yes
\end{itemize}

Overall, how satisfied are you with the interview?

\begin{itemize}
    \item Very dissatisfied
    \item Dissatisfied
    \item Neutral
    \item Satisfied
    \item Very satisfied
\end{itemize}

How well did the interviewer understand your responses?

\begin{itemize}
    \item Very poorly
    \item Poorly
    \item Neutral
    \item Well
    \item Very well
\end{itemize}

Was your interviewer a human being or an AI interviewer?

\begin{itemize}
    \item Human Interviewer
    \item AI Interviewer
\end{itemize}

If previous answer was “AI Interview”, then give the following questions:

\textbf{For AI Interviewer Group:}

How human-like did you find the AI interviewer's responses?

\begin{itemize}
    \item Not human-like at all
    \item Somewhat human-like
    \item Moderately human-like
    \item Very human-like
    \item Extremely human-like
\end{itemize}
Did you mainly use text or voice while being interviewed by the chat bot?

\begin{itemize}
    \item Mainly text
    \item Mainly voice
    \item Both text and voice
\end{itemize}

How well did the voice input work?

\begin{itemize}
    \item Did not try
    \item Tried. Voice input did not work at all
    \item Tried. Voice transcription was poor
    \item Tried. Voice Transcript was good  
\end{itemize}

\subsection{Interview responses: Example for thinking out loud}
\textbf{AI interviewer: }Given this context, how would you define the term "politics"?

\textbf{Respondent: }it's a pretty hard question to define the term politics I think for me politics is just the thing where you think about that Berlin and the German ambassadi and all the politicians and the all the how is it called all the parties and stuff like that also the election but not also it's not only Berlin it's also like really the politics also in the city of Munich for example I think politics is just a really poor thing and a lot of things are politics it starts with I don't know with the other universities stuff is a lot of politics money stuff it's a lot of politics and all the things I think it's it's a really wide term for politics at the end of the day for me politics such as all the rules and all the Decisions which are made for the complete people in Germany

\end{document}